%% file: cas-sc-template.tex
\documentclass[a4paper,fleqn]{cas-sc}

\usepackage{subcaption}
\usepackage{algorithm}
\usepackage{algpseudocode}
\usepackage{dashrule}

\usepackage[authoryear,longnamesfirst]{natbib}

\def\tsc#1{\csdef{#1}{\textsc{\lowercase{#1}}\xspace}}
\tsc{WGM}
\tsc{QE}

\begin{document}
\let\WriteBookmarks\relax
\def\floatpagepagefraction{1}
\def\textpagefraction{.001}

\shorttitle{}    

\shortauthors{}  

\title [mode = title]{Shrinkage‑Constrained Functional Calibration for Complex Computer Models}  

\tnotemark[1] 

\tnotetext[1]{} 

%

\author[1]{Liam Myhill}[orcid=0000-0002-0767-3799]



\ead{lmyhill@clemson.edu}


\credit{}

\affiliation[1]{organization={Clemson University Department of Mechanical Engineering},
            addressline={Fluor Daniel EIB}, 
            city={Clemson},
            postcode={29630}, 
            state={SC},
            country={United States}}

\author[1,2]{Enrique Martinez}[orcid=0000-0002-2690-2622]


\ead{enrique@clemson.edu}


\credit{}

\affiliation[2]{organization={Clemson University Department of Materials Science and Engineering},
            addressline={AMIC}, 
            city={Clemson},
            postcode={29630}, 
            state={SC},
            country={United States}}

\author[3]{Sez Russcher}
\ead{sez@clemson.edu}
\affiliation[3]{organization={Clemson University Department of Mechanical Engineering},
            addressline={Zucker Family Research Complex}, 
            city={Charleston},
            postcode={29401}, 
            state={SC},
            country={United States}}




\begin{abstract}
We propose a new Bayesian model calibration formalism as an alternative to the Kennedy – O’Hagan (KOH) framework which we term integrated bias with full uncertainty (IBFU). In KOH, calibration parameters are modeled as fixed, but unknown distributions with relatively weak prior constraints, and their posteriors are inferred jointly with an additive discrepancy Gaussian Process (GP). This formulation often provides limited regularization and leads to confounding pathologies when applied to inexact models with sparse, noisy measurements. By contrast, we represent each calibration parameter as the sum of a fixed best‑estimate value and a parameter correction represented by an independent GP over the input space, equipped with strong shrinkage priors. Any residual discrepancy that cannot be addressed via parameter correction is captured by an additive discrepancy GP operating on the simulator, similar to KOH. We then impose orthogonality constraints to mitigate confounding between the simulator and modeled additive discrepancy and co-linearity between model parameters. Imposing strong complexity shrinkage via conservative hyperpriors forces the mean parameter correction to remain flat across the domain, resulting in predictions that essentially converge with the KOH formulation. However, upon relaxing complexity shrinkage, should the data provide evidence that the effective calibration parameter varies across the domain, the mean parameter correction is allowed to become a function of the domain in a controlled, structured manner. In this sense, our approach is more universal: it effectively nests KOH as a special case while extending it to input‑dependent calibration, and it is more tightly constrained, because it anchors the true values around the best estimates and the shrinkage prior actively regularizes the calibration parameters.
\end{abstract}




\maketitle

\section{Introduction}

Calibrating inexact models with multiple sources of uncertainty, ontological, epistemic, and aleatory, requires careful consideration of the model inputs. These inputs consist of parameters which may have a physical basis, or some phenomenological fitting, and depending on the quality of available data, each may be poorly defined. In most practical model calibration problems, domain experts possess reasonably reliable best‑estimate values for the calibration parameters, often derived from physics, material specifications, engineering judgment, or previous experiments \cite{Cacuci01052010}. The standard Kennedy O’Hagan (KOH)\cite{Kennedy2001},  or subsequent expansions \cite{damianou_deep_2013,atamturktur_state-aware_2015,plumlee_bayesian_2017,plumlee_computer_2019}, treat these parameters as unknown distributions that need to be learned from the data constrained only with diffuse priors. This treatment does not fully leverage the best estimate information available to the modelers. As a result, the likelihood dominates the posterior, the best-estimate value is neglected as an anchor when inferring posterior distributions of each calibration parameter, which further contributes to the well‑known pathology of confounding between parameter calibration and discrepancy bias correction, even in modularized implementations \cite{on-the-fly-calibration}.

In calibration of uncertain, inexact models, one must consider the role of model discrepancy, and whether it can be attributed to physical parameters \cite{maupin_model_2020} or model inadequacy \cite{plumlee_bayesian_2017}. The interplay between multiple forms of uncertainty has lead to the so-called identifiability issue with KOH-style calibration \cite{paquette-rufiange_goal-oriented_2026}. There have been various attempts to address identifiability, including encoding structural information in the form of informative prior distributions, shape constrains, expected bounds, or derivative information \cite{semochkina_incorporating_2025}; adapting low-rank or basis reduced formulations \cite{bayesian_pca}, enforcing orthogonality constraints between the GP prior on discrepancy and the gradient of the computer model with respect to the calibration parameters \cite{grilches}, and modularization approaches that cut the feedback from discrepancy estimates to the calibration parameter ($\theta$) inference \cite{tohme_generalized_2020}. Early work can be traced to the work of Higdon \textit{et al} \cite{higdon_combining_2004,higdon_computer_2008}  and Bayarri \cite{bayarri_computer_2007,bayarri_framework_2007}, where methods such as Fourier series decomposition, principal component analysis, or wavelet decomposition are explored. If one is to account for both epistemic uncertainty in parameters, and the ontological uncertainty of model inadequacy using a Bayesian approach, then model priors play a crucial role in determining the extent to which these uncertainties interact to inform posterior predictions. These priors encode the degree of certainty that exists in expert opinion, and is commonly abandoned through the prescription of noninformative or weakly informative priors \cite{lemoine_moving_2019}.

We propose an alternative model calibration philosophy that preserves expert knowledge by anchoring each parameter at its best‑estimate value and allows deviations from that value through an additive correction. This correction could be a global or local offset represented either by a constant covariance (global) or state-dependent (local) functional correction modeled as a Gaussian Process (GP) across the domain. If one considers the more general case of state-dependent parameter variation, yet prescribes shrinkage hyperparameter priors that penalize GP mean deviations from zero, the resultant method is approximately equivalent to a model with global constant covariance. This formalism maintains parameter interpretability, and respects established domain knowledge on model parameters. Data‑driven departures from best-estimate values of calibration parameters ($\theta$) is permitted only when strongly supported by data. In this sense, the approach which is termed the integrated bias with full uncertainty (IBFU) is more flexible, yet better constrained and more faithful to the best-estimate information that modelers possess.


This paper is organized as follows: Section \ref{sec:ca-up} describes the development and implementation of the integrated bias approach and provides some context on shrinkage priors for optimal use cases of the method. Section \ref{sec:synthModel} describes academic examples that were specially engineered to thoroughly evaluate the method’s capabilities. Sections \ref{sec:results} and \ref{sec:discussion} contain the results and key points of discussion respectively, before concluding remarks in Section \ref{sec:conclusion}.

\section{Integrated Bias Formalism with Full Uncertainty (IBFU)}
\label{sec:ca-up}

The continuous KOH formalism of Bayesian model calibration with respect to observation data is provided in Equation \ref{eq:KOH}.

\begin{equation}
    y(x)=\eta(x,\theta)+\delta_\eta(x)+\varepsilon
    \label{eq:KOH}
\end{equation}
    
\noindent where $y$ is some experimental observation, $x$ are the control variables defining the operational and environmental state of the system, $\eta$ is a deterministic model simulator of the observed phenomenon, $\theta$  is a vector of input parameters to the model with \textit{a priori} uncertainty as to their true value, $\delta_\eta$ is an additive discrepancy term summed with $\eta$ to correct model deficiencies, and $\varepsilon$ is an additive noise term meant to account for observation error. Common practice is to assign a normal distribution to $\varepsilon$ such that $\varepsilon\sim\mathcal{N}(0,\sigma^2)$, where $\sigma^2$ denotes a conventional noise variance.

A major pitfall with KOH is the lack of identifiability due to how $\delta_{\eta}(x)$ interacts with $\eta(x,\theta)$, especially when the data lacks clear and discernible trends. When $\theta$ are weakly constrained, $\partial\eta/\partial\theta$ is small, and GP priors on $\delta_\eta$ are sufficiently flexible, one could potentially satisfy $\eta(x, \theta_1)+\delta_1(x) = \eta(x, \theta_2)+\delta_2(x) $ for all observed $x$, at which point ($\theta_1, \delta_1)$ and $(\theta_2, \delta_2)$ both become plausible. The interplay between epistemic $\theta$ uncertainty and ontological $\delta(x)$ uncertainty are the primary contributors to confounding, with aleatory contributions of $\varepsilon$ seen as secondary. In such scenarios, the deviations between the simulator and observations cannot be attributed – in a reliable manner – to parameter corrections or structural errors. 

In this paper, we re-imagine parameter calibration by concentrating the uncertainty mass near interpretable deviations from empirical best estimates $\theta^0$:

	\begin{equation}
	    \theta^\dagger=\theta^0+\kappa
        \label{eq:thetaDagger}
	\end{equation}

In Equation \ref{eq:thetaDagger}, the calibration parameters $\theta$ are reinterpreted as $\theta^\dagger$,  a summation of a fixed best-estimate value $\theta^0$ and a constant deviation from this estimate $\kappa$. However, many real systems violate the assumption that calibration parameters are global constants \cite{atamturktur_state-aware_2015}. In fact, model parameters sometimes act as \textit{effective parameters} , summarizing effects of unresolved physics instead of immutable physical properties \cite{brynjarsdottir_learning_2014}. In such scenarios, the additive correction on $\theta$ can be modeled as a state-dependent functional correction represented with a GP over the input space, 

 \begin{equation}
	    \theta^*=\theta^0+\kappa(x)
        \label{eq:thetaStar}
	\end{equation}
    
In Equation \ref{eq:thetaStar}, $\kappa(x)$ is a GP with shrinkage priors that, unless provided statistically significant reason to deviate, converge the mean towards zero. In this formalism, $\kappa(x)$ hyperparameters can be directly linked to tolerances, material variability, expert bounds, and modelers confidence on parameterizations. If all model inadequacy could be explained through reparameterization, then Equation \ref{eq:KOH} becomes

\begin{equation}
    y(x)=\eta(x,\theta^0+\kappa(x))+\varepsilon
    \label{eq:intDelta}
\end{equation}

In Equation \ref{eq:intDelta}, otherwise known as the integrated bias method, the $\{\theta,\delta_\eta\}$ confounding is structurally resolved. However, if further model inadequacy is expected to remain beyond reparameterization, the additive discrepancy term could be reintroduced albeit with increased risk of misattributing the disparity between model output and experimental observations.

\begin{equation}
y(x_i)=\eta(x,\theta^0+\kappa(x))+\delta_\eta(x)+\varepsilon
    \label{eq:ca-up}
\end{equation}
    
The novelty of the IBFU formalism shown in Equation \ref{eq:ca-up} stems from its preservation of best estimate values for calibration parameters and strong regularization of deviations from these best estimates via shrinkage priors \cite{ling_selection_2014,simpson_penalising_2015,yi_hierarchical_2012,piironen_hyperprior_nodate}. Shrinkage priors limit deviations that are not fully supported by data and  forces any corrections to the best estimates for both calibration parameters (i.e., $\kappa(x)$), and any corrections to the simulator prediction, (i.e., $\delta_\eta(x)$) to revert to zero, collapsing to the original best-estimate model, unless they are required by data. Hence, the proposed approach is more flexible as it can allow $\theta^*$ to vary across the domain in a controlled manner, yet remain arguably better constrained because the model predictions are anchored around reliable \footnote{see Section \ref{subsec:shrinkage}} best estimates for $\theta$ (i.e. $\theta^0)$. 

\begin{center}
\hrule height 0.8pt
\vspace{0.4em}
\captionof{algorithm}{Integrated Bias Calibration with Full Orthogonalized Uncertainty}
\label{alg:fuid_ortho}
\label{alg:fuid_ortho}

\vspace{0.4em}
\hrule height 0.8pt
\vspace{0.6em}

\begin{algorithmic}[1]

\Require Observations $\{(x_i,y_i)\}_{i=1}^{N_o}$, emulator $\eta(x,\theta)$, fixed $\theta^0$

\Statex \textbf{Preprocessing: Orthogonalization Setup}

\State Compute local emulator sensitivities:
\[
J(x_i)=\frac{\partial \eta(x_i,\theta)}{\partial \theta}
\]

\State Construct global Jacobian:
\[
J_{\mathrm{glob}}=
\begin{bmatrix}
J(x_1)^T \\
J(x_2)^T \\
\vdots \\
J(x_{N_o})^T
\end{bmatrix}
\]

\State Compute singular value decomposition:
\[
J_{\mathrm{glob}} = U S V^T
\]

\State Construct sensitivity operator $G=J_{glob}$

\State Construct projection operator:
\[
P_G = G(G^TG)^{-1}G^T
\]

\State Initialize transformed discrepancy fields$ ^\dagger$
\[
\kappa_z^{(0)}(x), \quad \delta_\eta^{(0)}(x) 
\]

\State Initialize GP hyperparameters
\[
\phi_\kappa := \{\ell_\kappa,\sigma_\kappa\},
\qquad
\phi_{\delta_\eta}:=
\{\ell_{\delta_\eta},\sigma_{\delta_\eta}\}
\]

\State Initialize noise variance $\sigma^2$

\For{$t = 1,\dots,N_{\mathrm{MCMC}}$}

    \Statex \textbf{Step 1: Update transformed embedded discrepancy $\kappa_z(x)$ using Metropolis-Hastings (MH) or preconditioned Crank-Nicolson (pCN)}

    \For{$k = 1,\dots,d_\theta$}

        \State Propose
        \[
        \kappa'_{z,k}(x)\sim
        q(\cdot\mid \kappa_{z,k}^{(t-1)}(x))
        \]

        \State Map discrepancy into orthogonal sensitivity basis:
        \[
        \kappa'(x)=V\kappa'_z(x)
        \]

        \State Compute effective parameter field:
        \[
        \theta^*(x_i)=\theta+\kappa'(x_i)
        \]

        \State Evaluate likelihood:
        \[
        y_i\sim
        \mathcal{N}
        \big(
        \eta(x_i,\theta^*(x_i))
        +(I-P_G)\delta_\eta^{(t-1)}(x_i),
        \sigma^2
        \big)
        \]

        \State Accept/reject using MH ratio

    \EndFor

    \Statex \textbf{Step 2: Update additive discrepancy $\delta_\eta(x)$ (Gibbs)}

    \State Compute residual:
    \[
    r_i=
    y_i-
    \eta(x_i,\theta+V\kappa_z^{(t)}(x_i))
    \]

    \State Sample:
    \[
    \delta_\eta^{(t)}(x)
    \sim
    \mathcal{N}(\mu_\eta,\Sigma_\eta)
    \]

    using GP conditional posterior

    \State Orthogonalize additive discrepancy:
    \[
    \delta_{\eta}^{\perp\,(t)}(x)
    =
    (I-P_G)\delta_\eta^{(t)}(x)
    \]

    \Statex \textbf{Step 3: Update noise variance $\sigma^2$ (Gibbs)}

    \State Compute residual:
    \[
    e_i=
    y_i-
    \eta(x_i,\theta+V\kappa_z^{(t)}(x_i))
    -
    \delta_{\eta}^{\perp\,(t)}(x_i)
    \]

    \State Sample:
    \[
    \sigma^2\sim
    \mathrm{Inv\mbox{-}Gamma}(a',b')
    \]

    \Statex \textbf{Step 4: Update GP hyperparameters}

    \State Update $\phi_\kappa$ using MH or pCN

    \State Update $\phi_{\delta_\eta}$ using MH or Gibbs if conjugate

\EndFor

\State \Return Posterior samples of
\[
\kappa(x)=V\kappa_z(x),\quad \delta_\eta^\perp(x),\quad \sigma^2
\]

\Statex \hspace{\algorithmicindent}
\footnotesize
$\dagger$ $(0)$ denotes empty-array or best-guess initialization.

\end{algorithmic}
\vspace{0.6em}
\hrule height 0.8pt
\end{center}

Because the prior heavily penalizes nonzero $\kappa(x)$, the posterior will strongly favor explanations where $\kappa(x)$ stays near zero unless the data provides clear evidence of input-dependent variation in $\theta^*$. This effectively pins $\theta^*$ near $\theta^0$ by default and only allows structured deviations only when necessary to explain the data. The premise of the method inherently assumes the admissibility of $\theta^0$, yet allows for controlled exploration of the model parameter space, and when necessary, even outside the prior range for $\theta$. Because both $\kappa(x)$ (the calibration parameter correction) and $\delta_\eta(x)$ (the model discrepancy) are under shrinkage, the integrated delta approach offers a regularized decomposition of model error. To utilize the IBFU methodology implemented in Python, see \cite{caked-up}.



\subsection{Shrinkage Priors: Prescription and Optimization}
\label{subsec:shrinkage}
IBFU allows stricter regularization but does not escape fundamental identifiability limits. Therefore, it must be used consciously with explicit sensitivity analysis on shrinkage priors and nominal values. Accordingly, a sensitivity analysis is incorporated into the procedure to determine the role of each latent $\kappa(x)$ field in model predictions and to reveal whether apparent corrections are data-driven or prior-driven. This is particularly important as the shrinkage priors are the main leverage utilized to maintain the generalizability of the IBFU methodology. While in KOH, $\theta$ is weakly constrained by its prior \cite{ling_selection_2014} and $\delta_\eta(x)$ is uninhibited from absorbing residual discrepancy uniformed by $\theta$ sensitivities, using IBFU, $\theta^0$ is fixed at best estimate values that are deemed reliable, and any deviation from $\theta^0$ must be expressed through $\kappa(x)$, which is strongly shrunk toward zero. $\theta^0$ should be evaluated based on published specifications and assimilation of data from laboratory experiments \cite{Cacuci01052010}, or in the absence thereof, evidence marshaling of expert opinions.

While the posterior inference is dominated by the observed  data, the shrinkage priors dictate the flexibility of each embedded $\kappa(x)$ GPs and the additive $\delta_\eta(x)$ GP to learn and differentiate the sources of model-form error.
The concept of utilizing hierarchical shrinkage to penalize exploration is well-established in the literature \cite{yi_hierarchical_2012}, and multiple implementations of shrinkage parameters exist for differing models. One of the earliest examples of shrinkage penalization comes from Lasso and Tibshirani \cite{tibshirani_regression_1996}, who developed the $L_1$ penalty and subsequent double-exponential prior distribution where a $\lambda$ tuning parameter controls the degree of penalization or shrinkage. In this work, the hierarchical shrinkage is applied through covariance hyperparameters for $\{\kappa(x),\delta_\eta(x)\}$, which will be referred to as $\{\phi_\kappa,\phi_{\delta_\eta}\}$ and each $\phi:=\{\ell,\sigma\}$ where $\ell$ denotes the correlation length describing each field’s sensitivity to the application domain, and $\sigma$ denotes the expected variance or amplitude of the anticipated field. One can consider $\ell$ as an effective complexity shrinkage and $\sigma$ as an effective magnitude shrinkage. The utilized prior on the additive discrepancy field $\delta_\eta(x)$ is multivariate Gaussian, such that 

\begin{equation}
    p(\delta_k|\ell,\sigma)\propto|K_\delta^{-1}|\exp{(-\frac{1}{2}\delta_k^TK_\delta^{-1}\delta_k)}
    \label{eq:multivariateGaussian}
\end{equation}
and $K_\delta$ is the covariance of the $\delta_\eta(x)$ GP constructed via a radial basis function (RBF) kernel. The main shrinkage mechanism is the quadratic form $\delta_k^TK_\delta^{-1}\delta_k$, which is analogous to a ridge regularization in a correlated function space. 

The hyperparameter priors utilized in this work take a lognormal distribution, explicitly written as

\begin{equation}
\begin{aligned}
p(\ell) =
\frac{1}{\ell\sigma_\ell\sqrt{2\pi}}
\exp\left[
-\frac{(\log\ell-\mu_\ell)^2}{2\sigma_\ell^2}
\right],
\quad
p(\sigma) =
\frac{1}{\sigma\sigma_\sigma\sqrt{2\pi}}
\exp\left[
-\frac{(\log\sigma-\mu_\sigma)^2}{2\sigma_\sigma^2}
\right]
\end{aligned}
\end{equation}
where $\log\ell\sim\mathcal{N}\{\mu_\ell,\sigma_\ell^2\}$ and $\log\sigma\sim\mathcal{N}\{\mu_\sigma,\sigma_\sigma^2\}$ are required inputs for both the $\kappa$ and $\delta_\eta$ GPs. Currently, the same hyperparameter priors are deployed for $\kappa(x)$ GPs corresponding to each of the $\theta$s, but modification of prior distributions is trivial. The hyperparameters $\{\phi_\kappa,\phi_{\delta_\eta}\}$ are also inputs denoting initial states of each Markov chain, and control the complexity and magnitude shrinkage of each GP. Increasing $\ell$ effectively reduces the curvature of the posterior predicted field, acting as a complexity shrinkage. Consequently, a decrease in $\ell$  will lead to a mean prediction with a higher relative curvature. The influence of $\sigma$ is intuitive in nature, such that increasing $\sigma$ will bias posterior predictive means of $\{\kappa(x),\delta_\eta(x)\}$ to have a larger magnitude and vice-versa, operating as a magnitude shrinkage. The Metropolis-Hastings update of $\{\phi_\kappa,\phi_{\delta_\eta}\}$ can be considered as a random walk in log-space, where the hyperparameter update is only accepted if deviation from the hyperprior shrinkage is supported by data. The discrepancy fields are individually modeled as Gaussian processes with stationary RBF covariance kernels, while the kernel length scales and variances are themselves treated as uncertain hyperparameters endowed with lognormal hyperpriors. Due to the joint inference of $\{\kappa,\eta,\delta_\eta\}$, adjustment of a single hyperprior will influence the joint prediction of each field. To demonstrate this, a sensitivity analysis of hyperprior prescription is shown later in Figure \ref{fig:sensitivity_analysis}.

Prescription of the shrinkage priors, both complexity shrinkage and magnitude shrinkage, indicate the preliminary level of engineering confidence one has in their computational model. The hierarchy of shrinkage starts from $\{\phi_\kappa,\phi_\delta\}$ and is propagated through the covariance matrices $\{K_\kappa,K_\delta\}$ into each $\{\kappa(x),\delta_\eta(x)\}$ field, finally resulting in the  posterior predictive distribution of $y(x)$. A sensitivity analysis of calibration predictions at various $\{\phi_\kappa,\phi_{\delta_\eta}\}$ detailing the hierarchical propagation through the discrepancy fields $\{\kappa(x),\delta_\eta(x)\}$ is provided in Section \ref{sec:discussion}.

It is recommended to first explore tightly constrained $\kappa(x)$ discrepancy fields with hyperparameters $\phi_\kappa \sim\{\ell \sim O(10),\sigma \sim O(0.1)\}$, where the magnitude of $\ell$ is provided as a percentage of the application domain. For example, prescribing $\ell_\kappa=100$ results in a completely flat $\kappa(x)$ prediction, analogous to a KOH method. Limiting the  flexibility of $\kappa(x)$ will allow each discrepancy field to explore the parameter space as a function of the application domain. The order of magnitude shrinkage applied to $\kappa(x)$ determines the exploratory range of the parameter correction. Sufficiently lenient shrinkage can allow it to explore values outside of the prior range of $\theta$, aiding in the discovery of optimal physical parameters when prior ranges are misattributed. It is recommended that users utilize sensitivity analysis to define optimal prior distributions that avoid over-fitting and absorbing what would otherwise be attributed to model-form error. 


\subsection{Orthogonality Constraints and Basis Representations}

Even with the use of shrinkage priors, a central challenge of the integrated bias formalism remains confronting the confounding between $\kappa(x)$ and $\delta_\eta(x)$.  The governing idea is that posterior predictions should be corrected by the input parameters ($\kappa(x)$), and that additional discrepancies ($\delta_\eta(x)$) should only be considered if the observed phenomenon cannot possibly be explained by the input parameters. Using methodologies developed by Plumlee \cite{plumlee_bayesian_2017,plumlee_computer_2019}, the additive discrepancy $\delta_\eta(x)$ is projected orthogonally to the directions of $\kappa(x)$ sensitivity to reasonably diminish the role of additive discrepancy in the methodology. Finally, potential intra-parameter confounding between $\kappa(x)$ corrections to calibration parameters on $\eta$ also need to be considered using methods proposed by Higdon \cite{higdon_combining_2004,higdon_computer_2008}. 

A Jacobian matrix is computed along x for each $\kappa(x)$ to elucidate the specific role of each individual $\theta$ parameter. A singular value decomposition then allows for the projection of $\kappa(x)$ along the principal directions in $\theta$-space such that each $\kappa(x)$ is orthogonally oriented compared to other $\kappa(x)$ fields. This projection is computed at each MCMC iteration and the projection utilized to compute residual discrepancy. The implementation of PCA \cite{bayesian_pca,plumlee_bayesian_2017} effectively reduces the degree of intra-parameter confounding relative to the method routines, which do not aim to differentiate between the contributions of the various $\kappa(x)$ fields. The concept of a local sensitivity analysis elucidating $\theta$ contributions is established \cite{hill_local_2010,Constantine2015ActiveS}. 

To compute the local sensitivity of the emulator to each $\kappa(x)$, the Jacobian is computed at all observation locations centered about $\theta^0$:

\begin{equation}
    J(x)=\frac{\partial\eta(x,\theta)}{\partial\theta}\in \mathbb{R}^{d_\theta}
\end{equation}
\label{eq:localJacobian}
where $\eta$ is the prior emulator. The global Jacobian then takes the form:

\begin{equation}
    J_{glob}=\begin{bmatrix} J(x_1)^T \\ J(x_2)^T \\ ... \\ J(x_{N_{obs}})^T \end{bmatrix}\in \mathbb{R}^{N_{obs}\times d_\theta}
\label{eq:globalJacobian}
\end{equation}
$J_{glob}$ is then decomposed according to Equation \ref{eq:jacobianDecomp}

\begin{equation}
    J_{glob}=USV^T
\label{eq:jacobianDecomp}
\end{equation}
where $V\in\mathbb{R}^{d_\theta\times d_\theta}$ is the orthogonal basis in parameter space and the columns of $V$ indicate the principal directions of sensitivity. The $\kappa(x_i)$ fields can then be mapped in the orthogonal basis to align with the sensitive directions of the GP emulator:

\begin{equation}
    \kappa(x)=V\kappa_z(x)
    \label{eq:jacobinMapping}
\end{equation}
where $\kappa_z(x)$ is the parameter discrepancy field in the non-transformed basis. 

To project the additive discrepancy orthogonal to the sensitivities of each $\kappa(x)$, a sensitivity operator $G$ is defined such that $G=J_{glob}$, then utilized to construct a projection operator $P_G$ which takes the form:

\begin{equation}
    P_G=G(G^TG)^{-1}G^T
\label{eq:projOp}
\end{equation}

The orthogonal component of the additive discrepancy $\delta^{\perp}_\eta(x_i)$ is then solved by Equation \ref{eq:orthoDelta}

\begin{equation}
    \delta^{\perp}_{\eta}(x)=(I-P_G)\delta_\eta(x)
    \label{eq:orthoDelta}
\end{equation}
which enforces the relation $G^T\delta^{\perp}_{\eta}(x)=0$.

The orthogonalized calibration methodology is then summarized by Equation \ref{eq:fullOrtho}

\begin{equation}
    y(x)=\eta(x,\theta^0+V\kappa_z(x))+(I-P_G)\delta_\eta(x)+\varepsilon
    \label{eq:fullOrtho}
\end{equation}

Implementing orthogonality constraints significantly improves IBFU, both in reducing the confounding of input parameter correction $\kappa(x)$, and additive model-form discrepancies $\delta_\eta(x)$. Whereas the prescription of the defined shrinkage priors $\{\phi_\kappa,\phi_{\delta_\eta}\}$ ensures that data supports deviation from best-estimate values, orthogonality constraints further ensure that said deviations only occur in directions of model sensitivity.



\section{Conceptual Problem Definition}
\label{sec:synthModel}

To benchmark  IBFU against conventional calibration methods and IB, a synthetic problem is conceived to generate representative observation data, with known true values of $\theta$ parameters. Simulation data is then generated, by sampling a uniform prior distribution of $\theta$ parameters. Equation \ref{eq:toyProb} describes the form of the observed data, where known values $\{\theta_1(x),\theta_2(x),\delta_\eta(x)\}$ could be compared against each method’s posterior predictions of each corresponding field. Two studies are conducted using the synthetic problem: (1) where the true  $\theta$ parameters are constant, and (2) where the true $\theta$ parameters vary along the application domain.  

The number of synthetic observation data-points $N_{obs} = 100$ and synthetic simulation data-points $N_{sim} = 500$ are intended to be representative of real-world cases where simulation data is more readily available than experimental observation. Table \ref{tab:trials} outlines the true $\{\theta,\delta_\eta(x)\}$ values used to create the synthetic observation dataset. For brevity, the  shrinkage priors and lognormal hyperparameter distribution used to generate the full results in Section \ref{sec:results} is also contained in Table \ref{tab:trials}. 


In each case, a known $\delta_\eta(x)=\cos(x)$ is added to the functional form of the observations, yet withheld from the functional form of the generated simulation data to control the systemic-ontological model discrepancy. The challenge then lies in the method’s ability to differentiate contributions of each $\kappa(x)$ from the additive discrepancy $\delta_\eta(x)$. The form $\cos(x)$ was specifically chosen as the $\delta_\eta(x)$ GP, because $\delta_\eta(x)$ is defined to have zero-mean, and is therefore best suited to handle models which do not consistently under or over predict the observed trends. Limitations such as this are elaborated on in Section \ref{subsec:limits}.

When generating simulation data, each $\theta$ is sampled via a latin-hypercube method from a uniform distribution $\theta\in[0,2]$ to imitate conventional sampling uncertain calibration parameters. In all study cases, the mean of the sampled $\theta$ is utilized as $\theta^0$ (i.e. $\theta^0=1$ for all $\theta$).  

\begin{figure}
    \centering
    \begin{subfigure}[t]{0.45\textwidth}
        \centering
        \includegraphics[width=\linewidth]{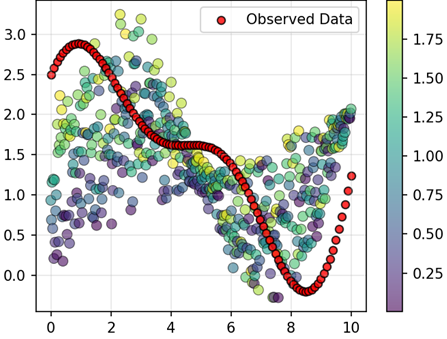} 
        \caption{Observation and simulation data (colored by $\theta_1$) of Study No. 1} 
    \end{subfigure}
    \hfill
    \begin{subfigure}[t]{0.45\textwidth}
        \centering
        \includegraphics[width=\linewidth]{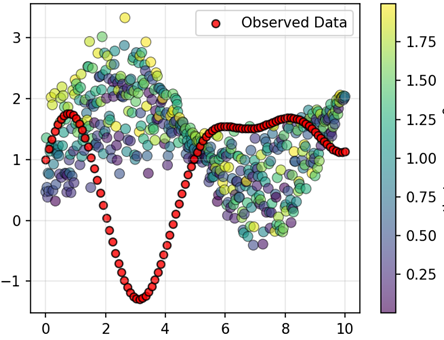} 
        \caption{Observation and simulation data (colored by $\theta_1$) of Study No. 2} 
    \end{subfigure}
\caption{Datasets for Studies No. 1 and 2}
\label{fig:data}
\end{figure}


The synthetic model combines a steady state response with a transient response and a linear drift taking the form
\begin{equation}
y(x)=\theta_1 e^{-0.3x}+\theta_2*\sin(\frac{2\pi x}{10})+0.2x+\delta_\eta(x)
    \label{eq:toyProb}
\end{equation}
where $\theta_1$ controls the amplitude decay through the exponential term, $\theta_2$ controls the oscillatory shape through the trigonometric sin function, and a linear trend is observed throughout. When generating synthetic simulation data, the $\delta_\eta(x)$ term is purposely omitted to produce the known true model-form error. 


\newcolumntype{C}[1]{>{\centering\arraybackslash}m{#1}}

\begin{table}[h!]
\centering
\renewcommand{\arraystretch}{1.35}
\setlength{\tabcolsep}{5pt}

\begin{tabular}{
|C{2.3cm}
||C{1.7cm}|C{1.7cm}|C{1.8cm}
||C{1.3cm}|C{1.3cm}|C{1.3cm}|C{1.3cm}|
}
\hline


& \multicolumn{3}{c||}{\textbf{True Problem Parameters}}
& \multicolumn{4}{c|}{\textbf{Utilized Shrinkage Hyperparameters}} \\
\hline


\textbf{Study No.}
& $\boldsymbol{\theta_1(x)}$
& $\boldsymbol{\theta_2(x)}$
& $\boldsymbol{\delta_\eta(x)}$
& $\boldsymbol{\ell_\kappa}$
& $\boldsymbol{\sigma_\kappa}$
& $\boldsymbol{\ell_\delta}$
& $\boldsymbol{\sigma_\delta}$ \\
\hline


1
& 1.5
& 1.75
& \multirow{2}{*}{$\cos(x)$}
& 10
& 0.1
& 0.1
& 0.1 \\
\cline{1-3}\cline{5-8}


2
& $\sin(x)$
& $\cos(x)$
&
& 0.5
& 0.2
& 0.1
& 0.1 \\
\hline


MD--DDD
& --
& --
& --
& 10
& 0.1
& 0.1
& 0.1 \\
\hline


\multicolumn{8}{|p{15.7cm}|}{
\textbf{Hyperprior Distributions:}

\centering
$\ell_\kappa,\ell_\delta
\sim
\mathcal{LN}(\mu=0.3,\sigma=0.5),
\qquad
\sigma_\kappa^2,\sigma_\delta^2
\sim
\mathcal{LN}(\mu=0.05,\sigma=0.5)$
\vspace{0.05cm}
} \\
\hline

\end{tabular}

\caption{
Study parameter configurations, utilized shrinkage hyperparameters,
and hyperprior distributions for the conceptual studies
and MD--DDD crystal plasticity application.
}

\label{tab:trials}

\end{table}

\section{Results}
\label{sec:results}

The results of calibration routines for each of the conceptual problems is presented using IBFU, the KOH method \cite{gattiker_combining_2006, gattiker_gaussian_2015} and the Integrated Bias (IB) method which neglects the role of additive discrepancy \cite{myhill2026bayesianmodelcalibrationintegrated}. The IB method has the same form as Equation \ref{eq:intDelta}. 


\subsection{Study No. 1}

\begin{figure}
    \centering
    \includegraphics[width=0.85\linewidth]{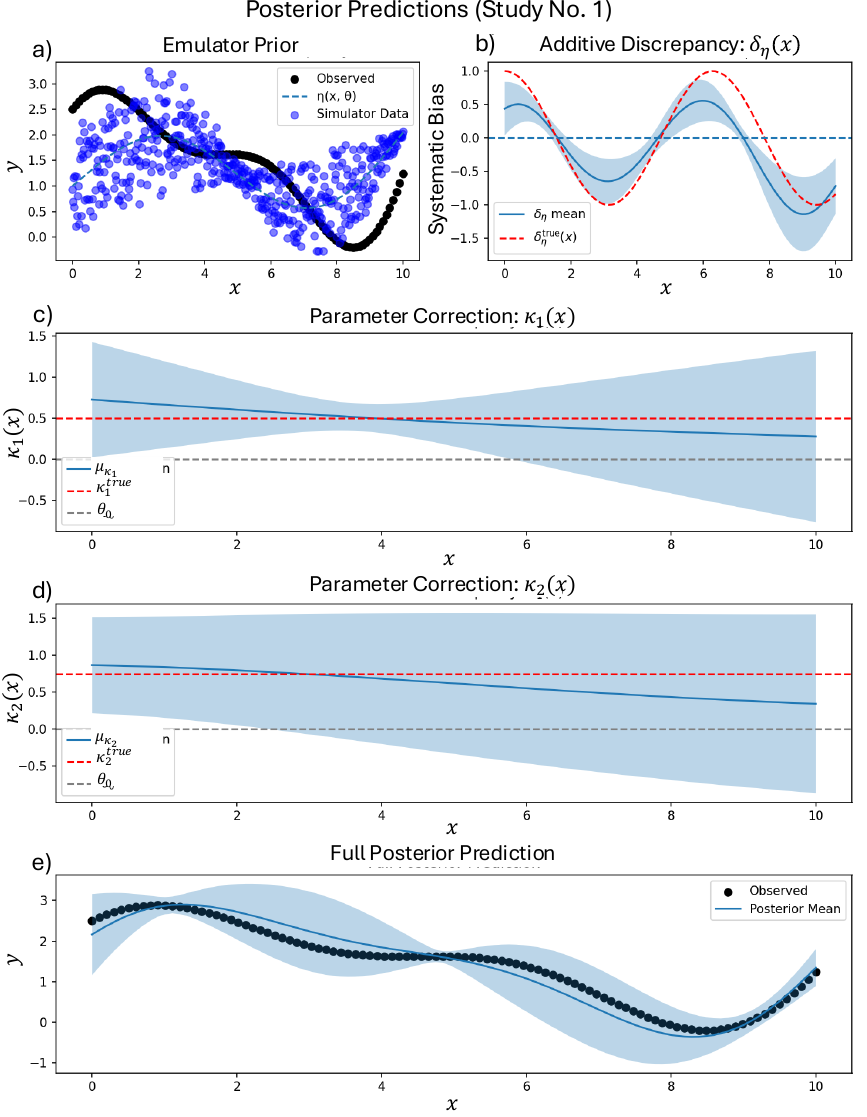}
    \caption{IBFU posterior predictions of Study No. 1}
    \label{fig:ca-up-trialID1}
\end{figure}

Study No. 1 is, counterintuitively, the more complex of the two study problems to solve using  IBFU method due to the constant nature of each $\kappa(x)$. Utilizing a $\kappa(x)$ GP leads to a degenerate case where the complexity shrinkage $\ell_\kappa\rightarrow\infty$ is the valid solution. This degeneracy can lead to a significant confounding without heavily constricting the shrinkage priors $\phi_\kappa:=\{\ell,\sigma\}$.  The utilized priors (from Table \ref{tab:trials}) allow for an emphasized flexibility of the $\delta_\eta(x)$ GP, with constriction of each $\kappa(x)$ GP. Prescribing larger values of $\ell_\kappa$ is necessary to handle the so-called degeneracy of Study No. 1, where the offset imposed by the mean of $\kappa(x)$ is simply a constant value. The posterior predictive results of Study No. 1 is given in Figure \ref{fig:ca-up-trialID1}. Each of the respective true values for $\{\kappa_1 (x),\kappa_2 (x)\}$ is contained within the posterior uncertainties, although the mean prediction does deviate from the ‘true’ values of each field. The same is true for the $\delta_\eta(x)$ GP, which captures the $\cos(x)$ functional form of the additive discrepancy, albeit with some deviations. The quality of the predictions may improve with an increased number of MCMC draws; however, the utilized value of $N_{MCMC}=20000$, which takes on average $\approx$ 2.5 hours to conduct on a single CPU, for the given amount of data, is deemed appropriate for demonstrating the capabilities of IBFU.

Figure \ref{fig:ca-up-trialID1} (a) shows the observation-simulation data along with the emulator prior. The additive discrepancy contributions to the posterior prediction is given in Figure \ref{fig:ca-up-trialID1} (b), while panes (c)-(d) show parameter correction fields differentiated by the input parameter. Finally, Figure \ref{fig:ca-up-trialID1} (e) demonstrates the full posterior predictions compared with the observation data. The same formatting is used for all Figures depicting IBFU posterior results, with varying numbers of $\kappa(x)$ fields depending on the number of identified calibration parameters. 

\begin{figure}
    \centering
    \begin{subfigure}[t]{0.42\textwidth}
        \centering
        \includegraphics[width=\linewidth]{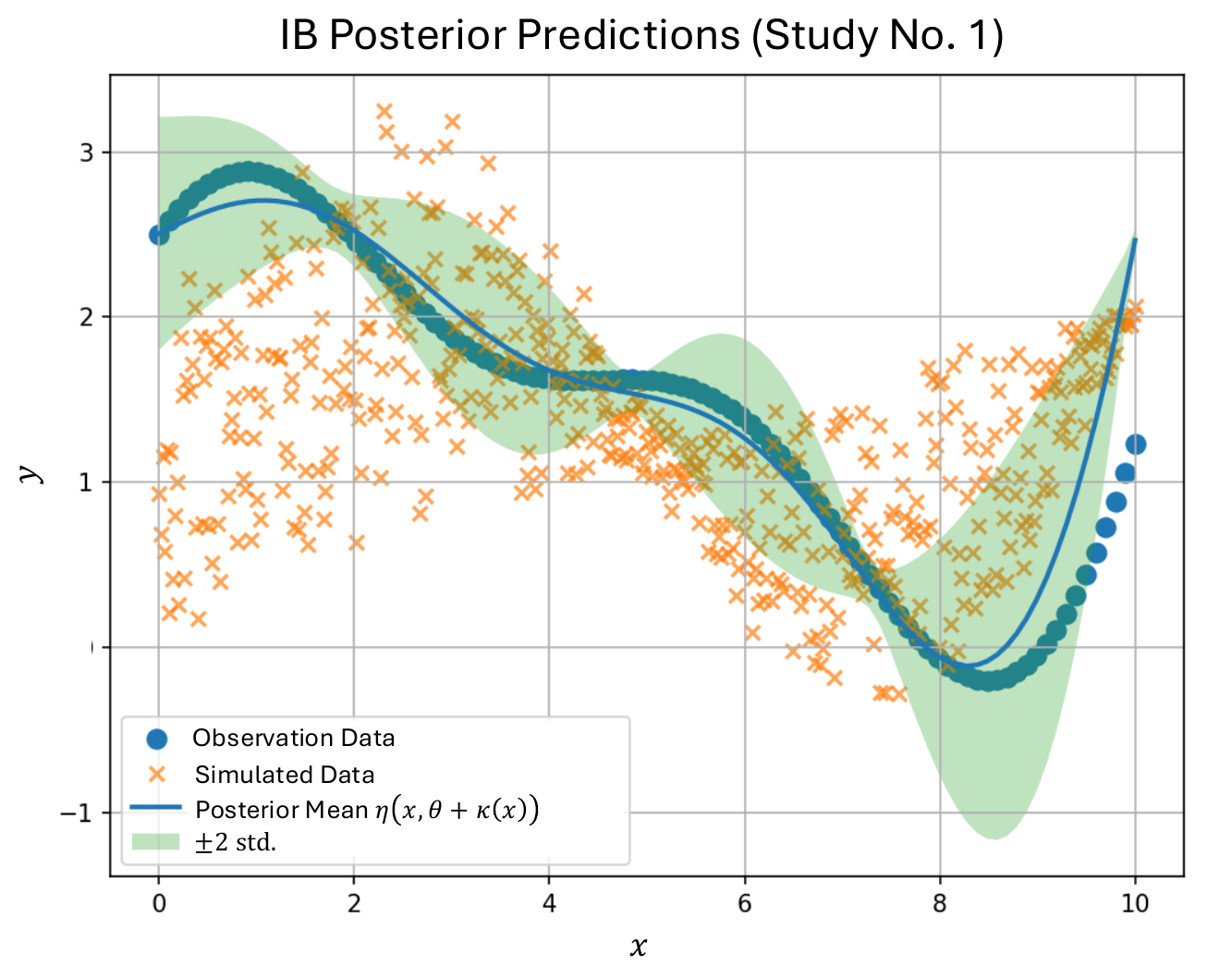} 
        \caption{IB posterior predictions of Study No. 1} 
    \end{subfigure}
    \hfill
    \begin{subfigure}[t]{0.49\textwidth}
        \centering
        \includegraphics[width=\linewidth]{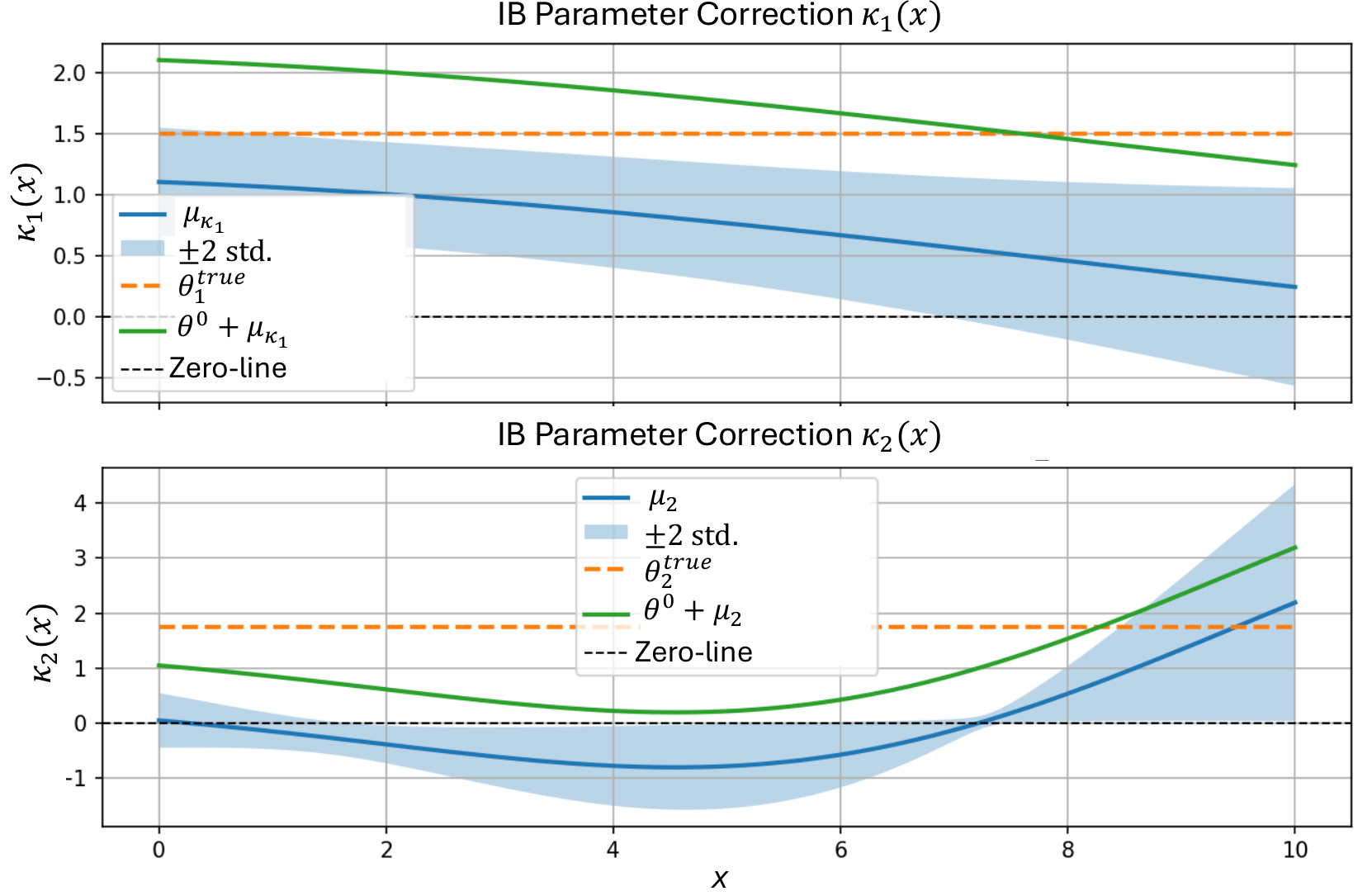} 
        \caption{Parameter correction predicted by IB} 
    \end{subfigure}
\caption{Integrated bias results of Study No. 1}
\label{fig:intDelta_trialID1}
\end{figure}

Figure \ref{fig:intDelta_trialID1} demonstrates the ability of the IB methodology to learn the observed trends of Study No. 1. Applying IB to Study No. 1 produces posterior predictions which match observations for the majority of the application domain, yet diverge significantly in the region $x\in[9,10].$ The uncertainty band for $\kappa_1(x)$, controlling amplitude decay significantly widens in this domain, indicating that the method cannot identify an input parameter that, when modified, enforces the GP to converge with the observation data. At this point, the modifications of the input parameters are not able to account for the additional external discrepancy field, which does not exist in the IB methodology. 

\begin{figure}
    \centering
    \begin{subfigure}[t]{0.55\textwidth}
        \centering
        \includegraphics[width=\linewidth]{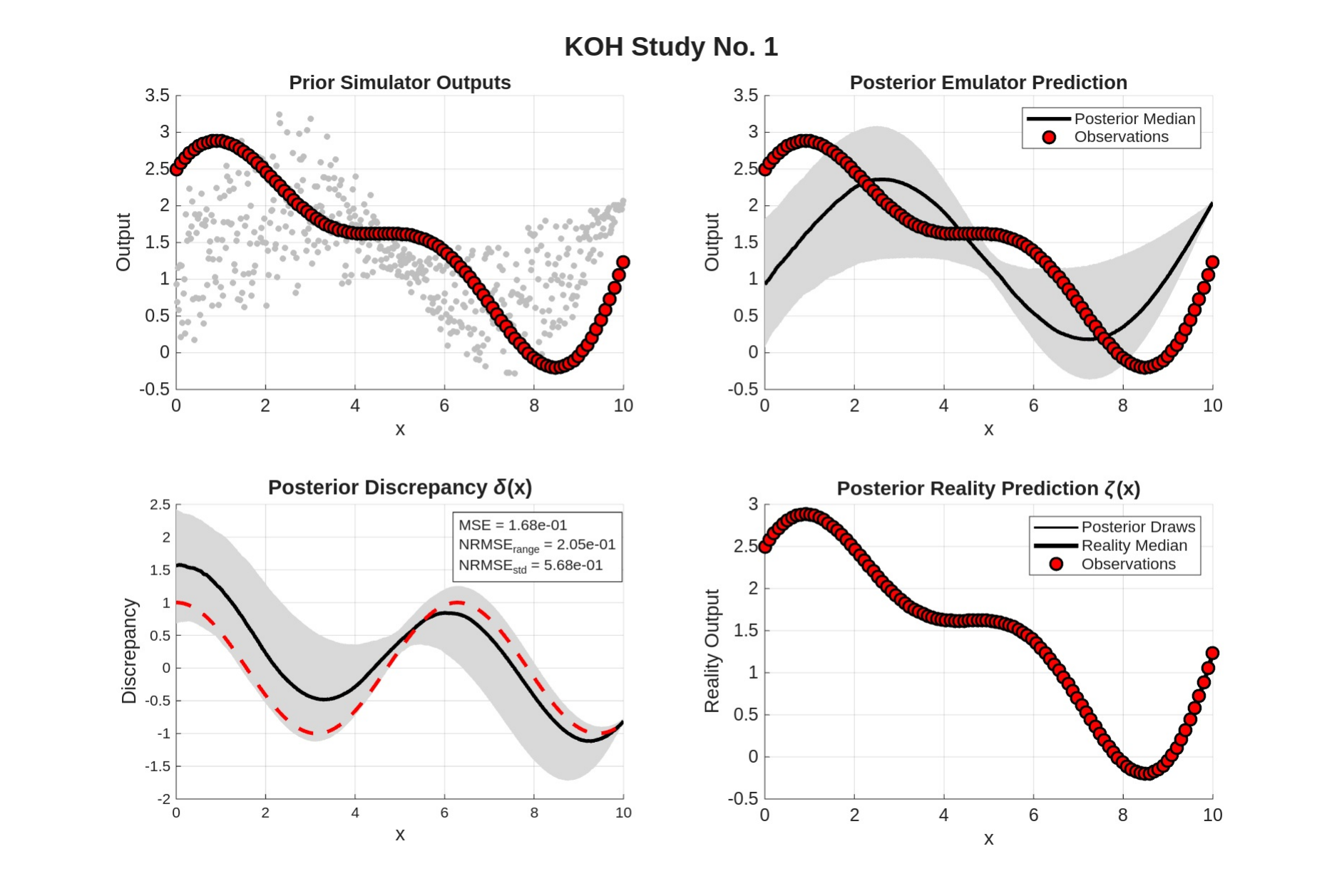} 
        \caption{KOH posterior predictions of Study No. 1} 
    \end{subfigure}
    \hfill
    \begin{subfigure}[t]{0.40\textwidth}
        \centering
        \includegraphics[width=\linewidth]{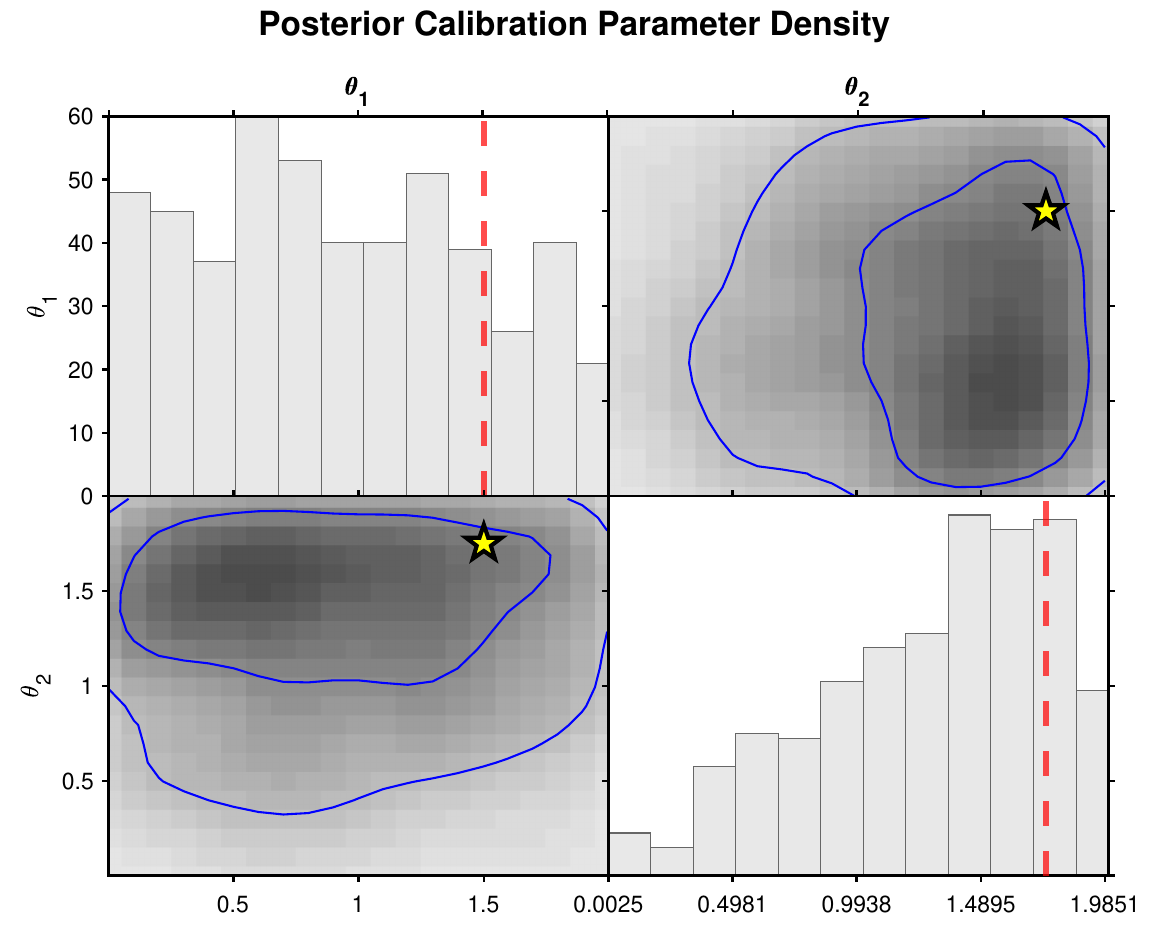} 
        \caption{Posterior parameter distribution output by KOH, where stars and red dashed lines denote true values} 
    \end{subfigure}
\caption{KOH results of Study No. 1}
\label{fig:gpmsa_trialID1}
\end{figure}

The results provided in Figure \ref{fig:gpmsa_trialID1} demonstrate KOH’s ability to learn the correction to each $\theta$ parameter, and the additive discrepancy field $\delta_\eta(x)$. KOH results for the dataset given in Study No. 1 predicts results which are over-fitted to the synthetic observation data and the posterior distributions for $\{\theta_1,\theta_2 \}$ lack both precision and accuracy, indicating that the distributions have converged without confidence. Ignoring the misparameterization, all the model-form error is then absorbed by $\delta_\eta(x)$. The artificially low uncertainty in the posterior predictions is evidence of the overfitting, whereby parameter uncertainty and additive discrepancy compensate for one another. Overfitting is not atypical for KOH-style calibration routines, and is well documented  \cite{identifiability-review}.

\subsection{Study No. 2}

\begin{figure}
    \centering
    \includegraphics[width=0.85\linewidth]{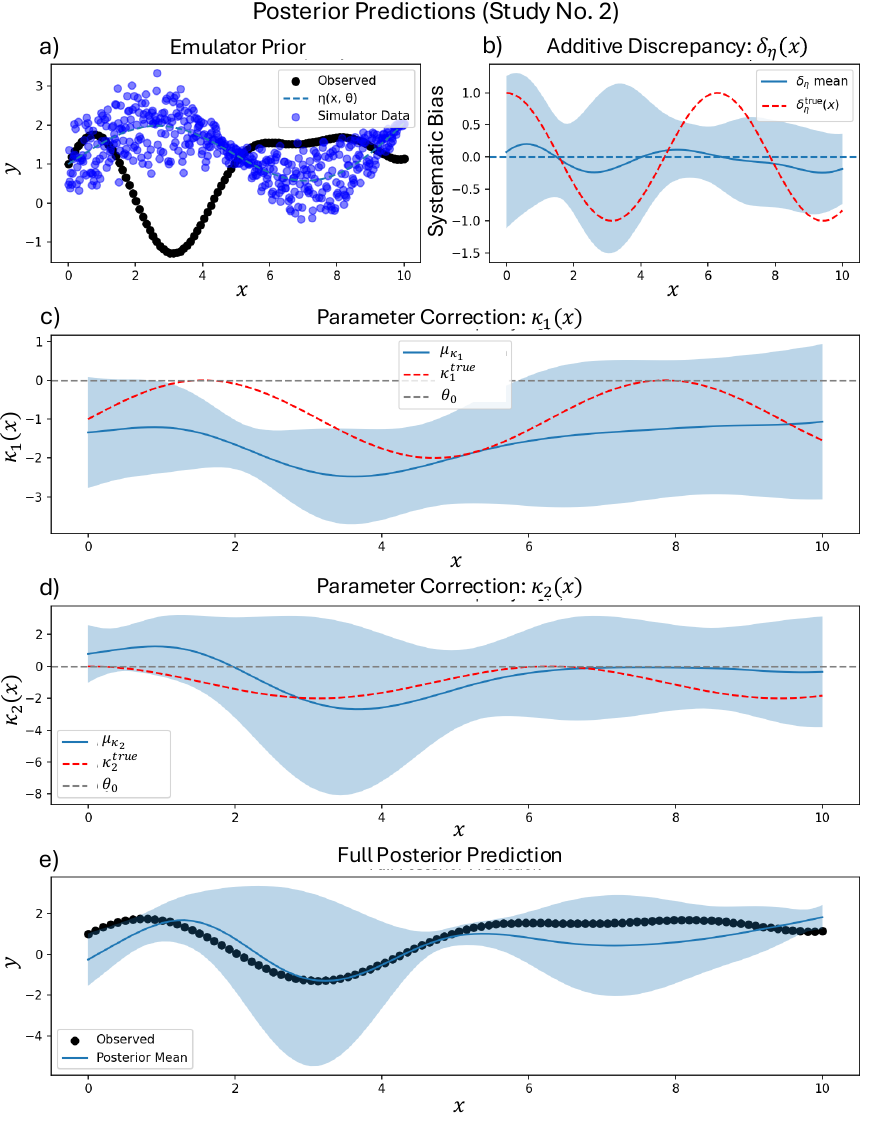}
    \caption{Results of IBFU applied to Study No. 2}
    \label{fig:ca-up-trialID2}
\end{figure}

One of the most attractive features of IBFU is its ability to learn corrections of input parameters which can evolve as a function of the application domain. Figure \ref{fig:ca-up-trialID2} demonstrates the ability of IBFU to simultaneously decouple parameter correction from additive discrepancy and treat each as functions of $x$. The shrinkage priors applied to generate this results set are given in Table \ref{tab:trials}. Further refinement of the hyperpriors may generate a set of results in better agreement with observation data; however, the utilized hyperpriors were chosen to be comparable with the hyperpriors of Study No. 1. 

\begin{figure}
    \centering
    \begin{subfigure}[t]{0.40\textwidth}
        \centering
        \includegraphics[width=\linewidth]{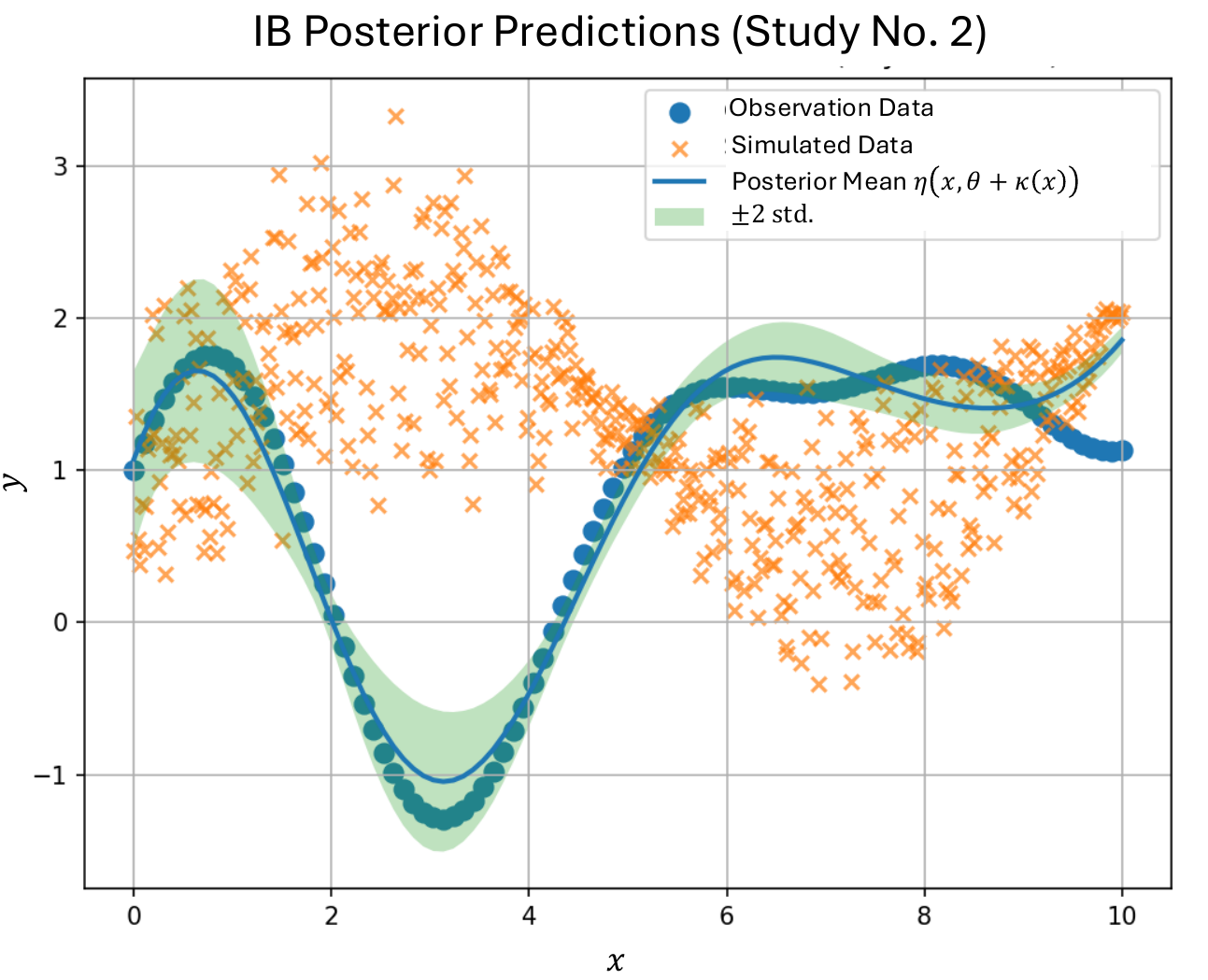} 
        \caption{IB posterior predictive of Study No. 2} 
    \end{subfigure}
    \hfill
    \begin{subfigure}[t]{0.51\textwidth}
        \centering
        \includegraphics[width=\linewidth]{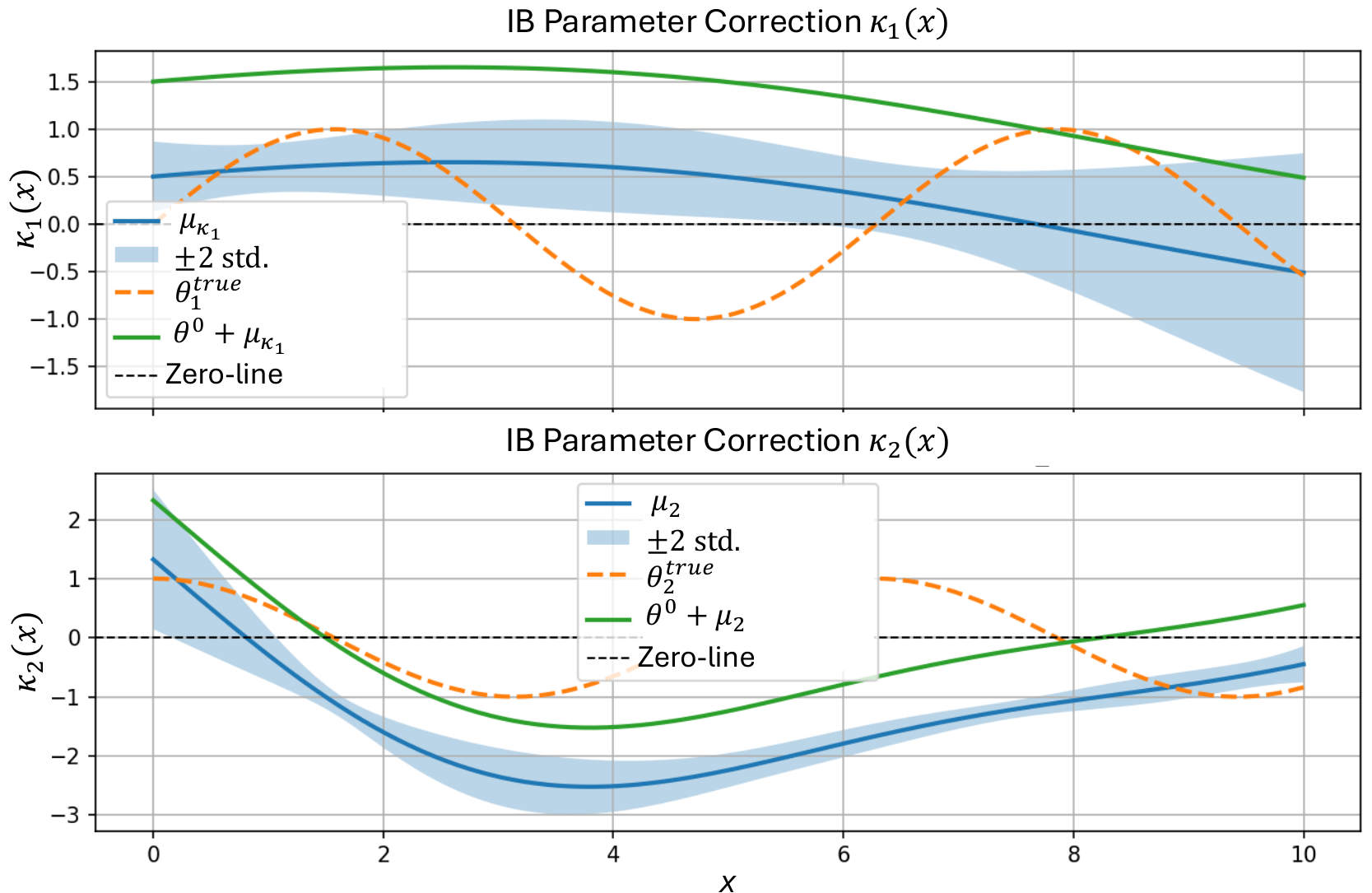} 
        \caption{Parameter correction predicted by IB} 
    \end{subfigure}
\caption{Integrated bias results of Study No. 2}
\label{fig:intDelta_trialID2}
\end{figure}

Figure \ref{fig:intDelta_trialID2} provides the IB results for Study No. 2, where the parameter correction is once again forced to account for a functional set of input parameters confounded with an additive discrepancy term that is absent from the IB formalism. As was the case in Study No. 1, the ID method struggles to produce a posterior prediction in agreement with observation data in $x\in[9,10]$. The reason is the same, due to the fact model inadequacy is not directly addressed.

\begin{figure}
    \centering
    \begin{subfigure}[t]{0.48\textwidth}
        \centering
        \includegraphics[width=\linewidth]{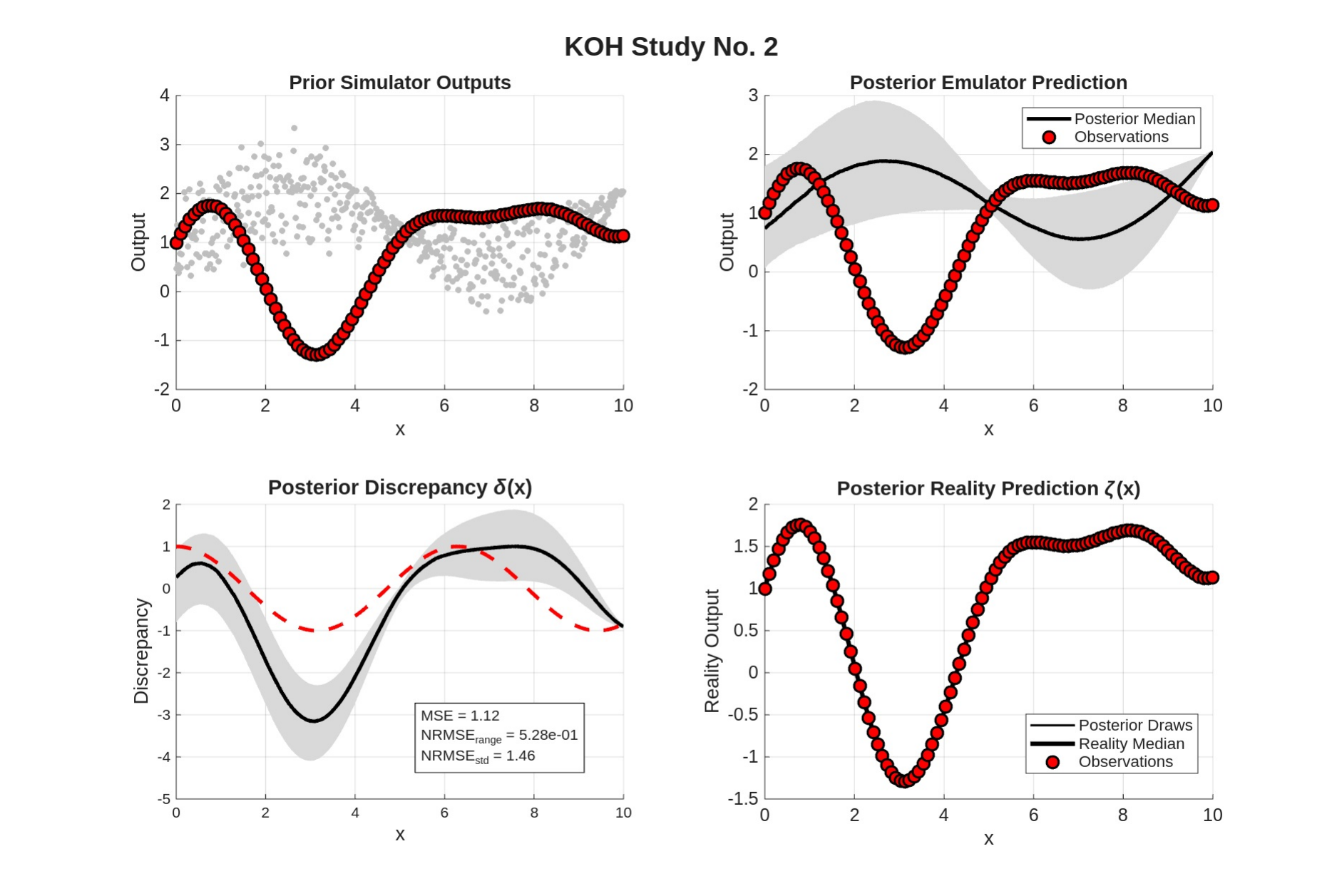} 
        \caption{KOH posterior predictions of Study No. 2} \label{fig:trialID1_gpmsa}
    \end{subfigure}
    \hfill
    \begin{subfigure}[t]{0.51\textwidth}
        \centering
        \includegraphics[width=\linewidth]{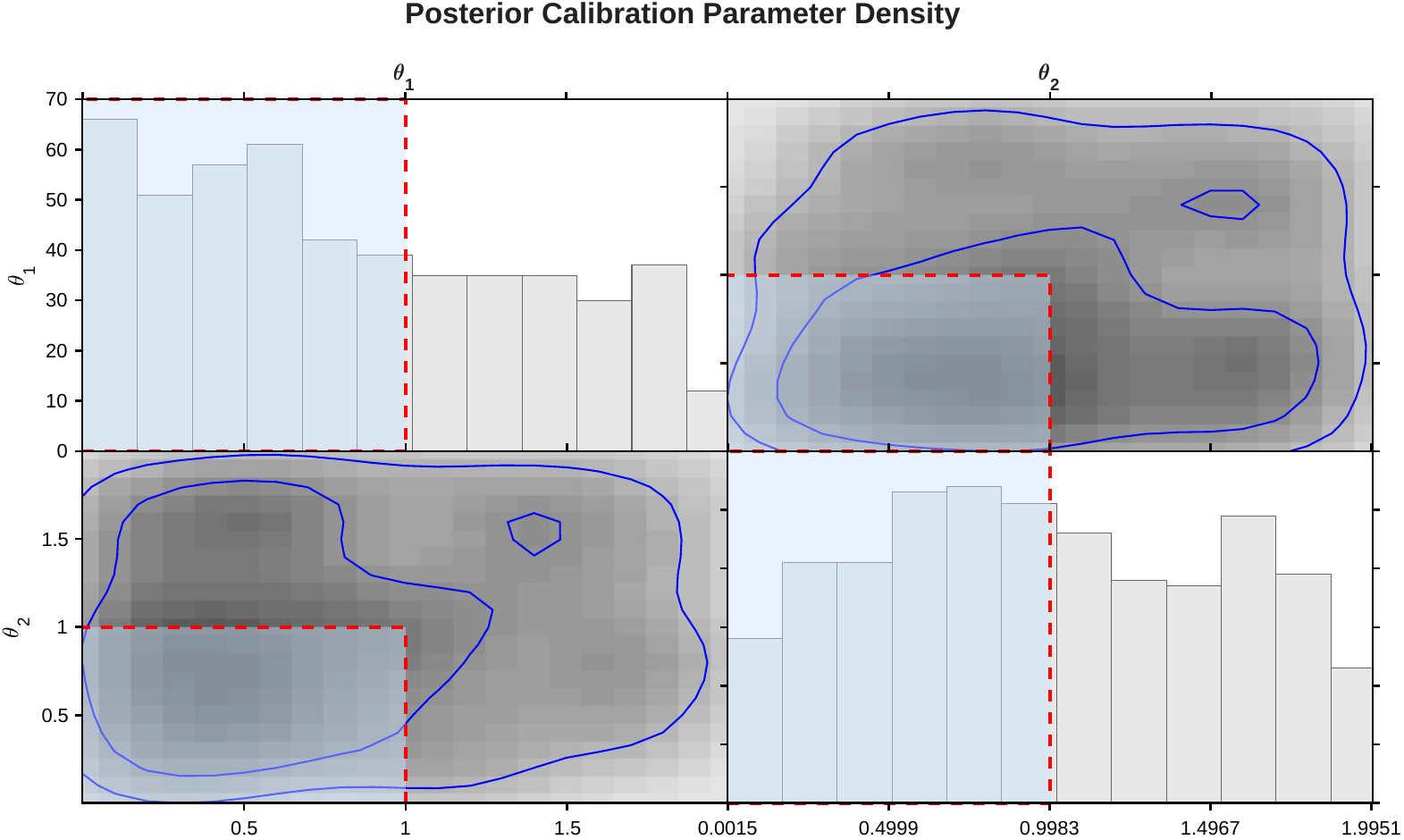} 
        \caption{Posterior parameter distribution output by KOH with highlighted true region} 
    \end{subfigure}
\caption{KOH results of Study No. 2}
\label{fig:gpmsa_trialID2}
\end{figure}

Figure \ref{fig:gpmsa_trialID2} shows the results of the KOH method for the dataset given in Study No. 2. The posterior distributions for each $\theta$ parameter are noisy due to the non-constant defined true $\theta$, as capturing functions of the application domain for $\theta$ is impossible with KOH. It is also worth noting the lack of predictive uncertainty, despite the uncertainty bands on the posterior emulator predictions and trained model discrepancy $\delta_\eta(x)$. As was the case for Study No. 1, the lack of uncertainty bands for the posterior reality predictions is verified to be another instance of overfitting.

\section{Calibration of Continuum Crystal Plasticity Models}

A real-world application is also presented, whereby IBFU is used to calibrate mesoscale defect models to atomistic simulation data, which is treated as experimental observation. Molecular Dynamics (MD) simulations feature simulation cells containing millions of degrees of freedom for length-scales in the range of nanometers, which is not scalable to bulk engineering materials. To remedy this, continuum simulation models of material behavior must be calibrated to capture the same trends observed in MD at the microscale, and extrapolate such behaviors to the meso-macroscale. When studying the plastic deformation of crystalline solids, Discrete Dislocation Dynamics (DDD) is the natural candidate model to facilitate the projection of crystal defect behavior to larger scales. 

In both MD and DDD, the critical resolved shear stress (CRSS) necessary to drive dislocations over local energy minima, $\tau^{CRSS}$, is quantified as a function of the lateral separation distance between dislocation dipoles. At or above $\tau^{CRSS}$, dislocations glide is uninhibited and plastic deformation of crystalline materials occurs. The input parameters to the DDD model known to influence the computed $\tau^{CRSS}$ are $\ell^C$, a dislocation core size prescribed in units of $\vec{b}$,  the shear modulus $\mu$ prescribed in units of Pa, and Poisson's ratio $\nu$, which is dimensionless.

\begin{figure}
    \centering
    \includegraphics[width=0.65\linewidth]{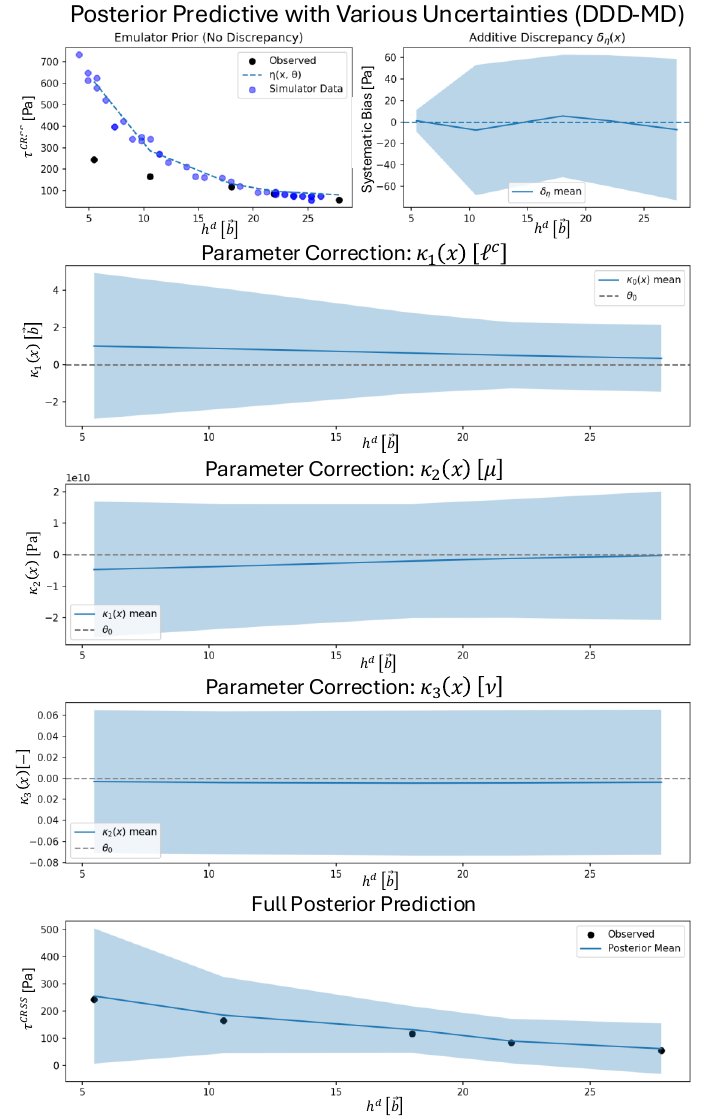}
    \caption{IBFU predictions of the various error sources in the calibration of continuum DDD model predictions of $\tau^{CRSS}$ to MD observations \cite{myhill2026bayesianmodelcalibrationintegrated}}
    \label{fig:ddd-md}
\end{figure}

Figure \ref{fig:ddd-md} demonstrates the calibrated prediction of $\tau^{CRSS}_{DDD}$ to observations made using equivalent MD simulation cells in Cu \cite{mishin_structural_2001} as a function of the separation distance between dislocation glide planes. A more detailed description of the problem statement is provided in \cite{myhill2026bayesianmodelcalibrationintegrated}, where the calibration is preformed using both the IB and KOH methods. For the best-estimate $\theta^0$ for the feature vector $\theta:=\{l^c,\mu,\nu\}$, the average of all sampled $\theta$ was used, identically to the conceptual problems and \cite{myhill2026bayesianmodelcalibrationintegrated}.

\begin{figure}
    \centering
    \includegraphics[width=0.75\linewidth]{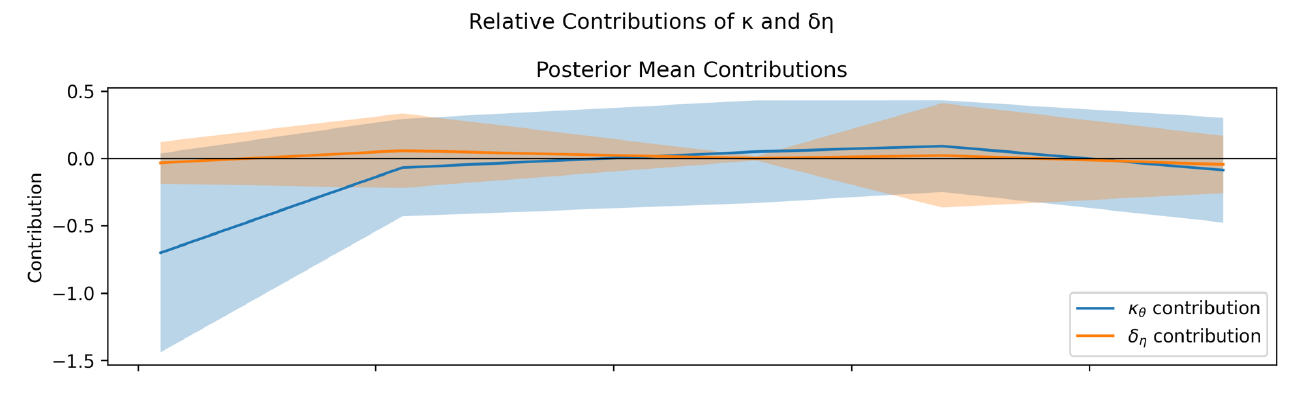}
    \caption{Relative contributions of $\kappa$ and $\delta_\eta$ GPs in generating posterior predictions of $\tau^{CRSS}$}
    \label{fig:relativeContribution}
\end{figure}

To decompose the effects of the $\kappa(x)$ corrections from the additive discrepancy, an analysis of the relative contributions to the posterior predictions is constructed via decomposition of $\eta(x,\theta^0+\kappa(x))+\delta_\eta(x)$ into constituent terms $\{\eta(x,\theta^0),\eta(x,\theta^0+\kappa(x)),\delta_\eta(x)\}$. The results of the analysis are given in Figure \ref{fig:relativeContribution}, and explicitly show the relative contributions of each discrepancy GP to the calibration of the DDD model. Figure \ref{fig:relativeContribution} demonstrates how $\kappa(x)$ dominates in the correction of $\tau^{CRSS}$, due to the sensitivity of the DDD model to input parameter $\mu$. The relative fractions are heuristic contribution measures and not strict variance decompositions, although the relative contributions of each field to the cumulative uncertainty is possible with IBFU. 

Application of IBFU to a practical calibration problem faced by the material science community demonstrates its ability to differentiate the various sources of error introduced when coarse-graining the multiplicity of physical mechanisms present at the atomistic scale. 
\section{Discussion}
\label{sec:discussion}

One significant challenge faced during the application of IBFU to the demonstrated conceptual problems was the selection of hyperpriors to capture the engineered behavior of the latent parameter discrepancy fields $\{\kappa_1 (x),\kappa_2 (x)\}$ and additive discrepancy $\delta_\eta(x)$. Over-optimizing the response for both studies is possible; however, in doing so the generalizability of the method is compromised. Without properly tailoring the shrinkage of hyperparameters $\{\phi_\kappa,\phi_{\delta_\eta}\}$, the inter-parameter confounding between $\kappa(x)$ and $\delta_\eta(x)$ may cause the fields to deviate significantly from the known values. 

To determine the degree to which shrinkage should be applied to the priors for the conceptual problems, a block-search algorithm was implemented which iterated over set permutations of the hyperpriors. Over 1000 combinations of the shrinkage priors were generated and each calibration routine was run in parallel using Clemson’s Palmetto high performance cluster \cite{antao2024modernizing}. The search was tailored to favor configurations with strong constraints on the $\{\kappa_1(x),\kappa_2(x)\}$ fields to mirror model calibration routines which assume constant input settings across the defined application domain. To capture the no-curvature, moderate-amplitude response necessitated for each  $\kappa$ discrepancy field, $\ell_\kappa$ ought to be relatively large $(O([10-100])$, and $\sigma_\kappa$ kept to $O(0.1)$ to remain proportional with the degree of confidence in the magnitude of model predictions. Supplementing these trends with a very flexible additive discrepancy field $\delta_\eta(x)$, (i.e. low $\ell_{\delta_\eta}$, moderate to large $\sigma_{\delta_\eta}$) can produce results akin to those found in Figure \ref{fig:ca-up-trialID1}. A dedicated sensitivity analysis of $\{\phi_\kappa,\phi_{\delta_\eta}\}$'s effects on $\{\kappa_1(x),\kappa_2(x),\delta_\eta(x)\}$ for both conceptual problems is provided in Figure \ref{fig:sensitivity_analysis}. 

It should be noted that real applications of IBFU will not have exact quantified values for input parameters $\kappa(x)$ nor definitive prior information on model inadequacies that necessitate the addition of $\delta_\eta(x)$. In such cases, the authors recommend first placing significant constraints on the flexibility of latent discrepancy fields $\kappa(x)$. These constraints should take the form of a complexity shrinkage on $\kappa(x)$ (increase in $\ell_\kappa$). Such constraints should only be relaxed if one can argue that parameter drift is a physically meaningful mechanism for addressing discrepancies. If it is known \textit{a-priori} that model predictions are far from experimental observations, increasing $\sigma_{\delta}$ will relax the magnitude shrinkage, and remove penalties for exploring larger magnitudes of $\delta_\eta(x)$. 

A demonstration of IBFU's tendency to minimize  parameter corrections from $\theta^0$ is made evident in the real-world example provided in Figure \ref{fig:ddd-md}, where both the additive discrepancy and calibration parameter correction are near-constant throughout the entire application domain. This result agrees strongly with those found in \cite{myhill2026bayesianmodelcalibrationintegrated}, where the shear modulus describing the elastic properties of single crystal Cu is identified by the data to be the most influential parameter in converging simulation results to observed data. The magnitude of the uncertainty band surrounding the $\delta_\eta(x)$ field never deviates beyond $\pm30\%$ of the predicted value, and the mean fluctuates at near-zero values, indicating that the model is capable of replicating observed data without the need for an additive correction, as was the central conclusion of \cite{myhill2026bayesianmodelcalibrationintegrated}. It is also worth noting the significant reduction in the amount of data available, with only 5 observation points and 43 simulation points used to learn the model sensitivities and dynamics.

\subsection{Sensitivity of the method to shrinkage priors}

To systematically evaluate how the priors affect the ability of the model to capture the desired trends, a loss function is constructed for the purpose of cross validation. The loss takes the form of Equation \ref{eq:calibrationLoss}. 

\begin{equation}
    loss=NRMSE(\eta|y_{obs})+NRMSE(\theta^0+\kappa(x_i)|\theta^{true})+NRMSE(\delta_\eta(x_i))|\delta_\eta^{true}(x_i))
    \label{eq:calibrationLoss}
\end{equation}
where $NRMSE$ describes the normalized root-mean squared error between known values and posterior predictive means, where normalization is conducted by dividing by the range of the signal.

\begin{figure}[h!]
    \centering

    \begin{subfigure}[t]{0.48\textwidth}
        \centering
        \includegraphics[width=\linewidth]{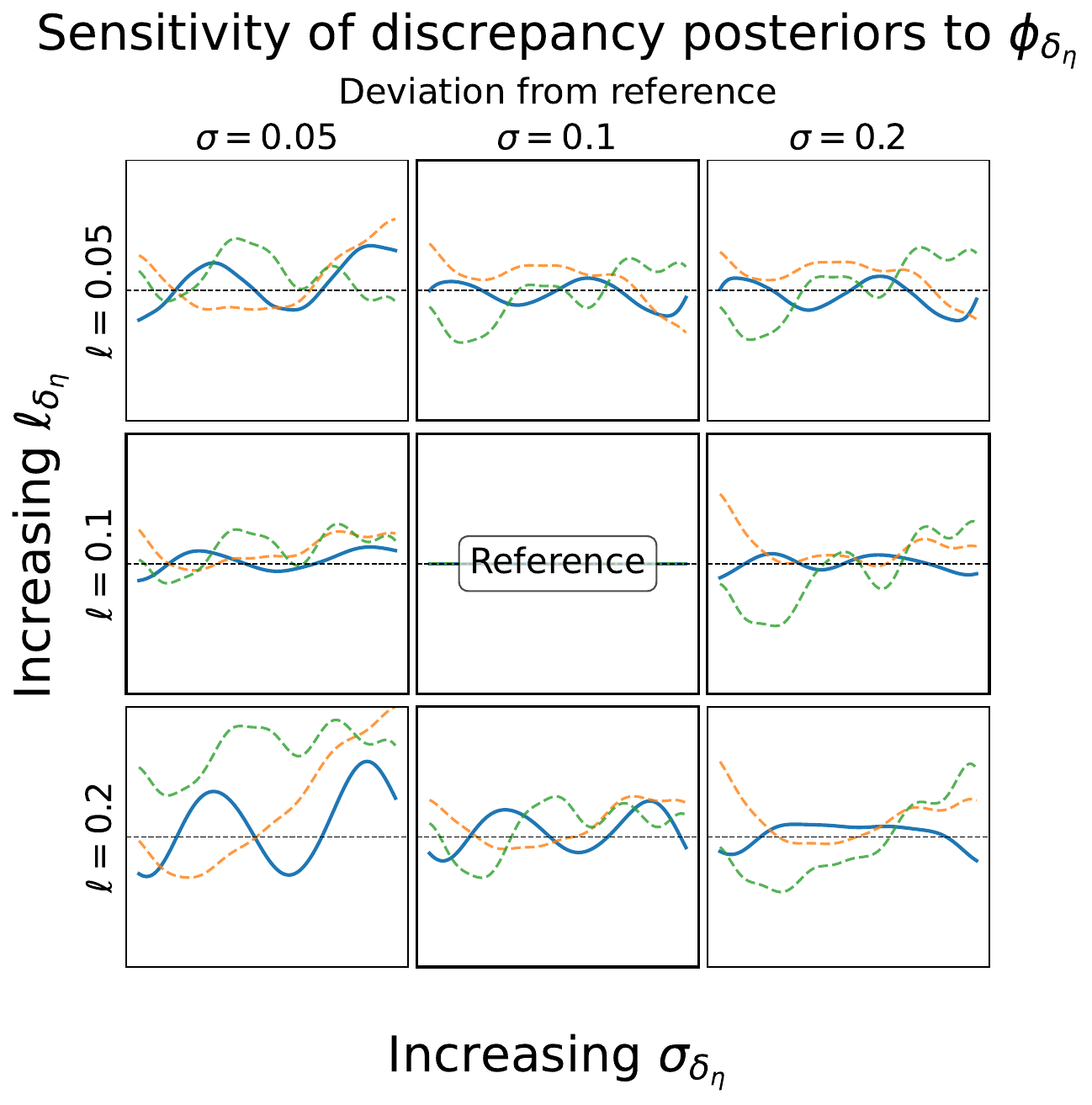}
        \caption{Sensitivity to $\phi_{\delta_\eta}$ for Study No. 1}
        \label{fig:sensitivity_e1}
    \end{subfigure}
    \hfill
    \begin{subfigure}[t]{0.48\textwidth}
        \centering
        \includegraphics[width=\linewidth]{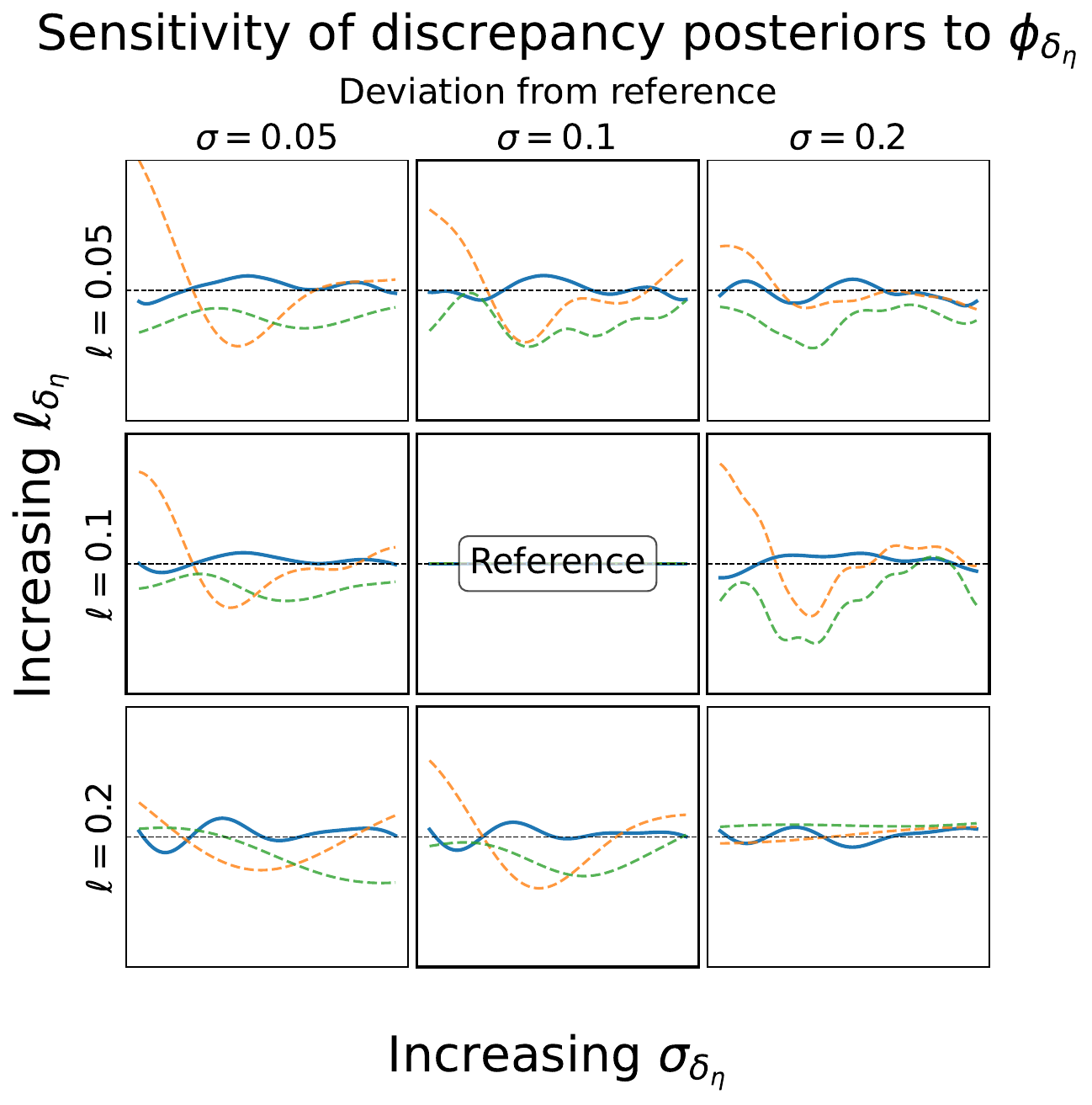}
        \caption{Sensitivity to $\phi_{\delta_\eta}$ for Study No. 2}
        \label{fig:sensitivity_e2}
    \end{subfigure}

    \vspace{0.4cm}

    \begin{minipage}{0.9\textwidth}
        \centering
        \footnotesize
        \begin{tabular}{ccc}
            \textcolor{blue}{\rule{1.5em}{2pt}} $\delta_\eta(x)$
            &
            \textcolor{orange}{\hdashrule[0.5ex]{1.5em}{2pt}{2pt 1pt}} $\kappa_1(x)$
            &
            \textcolor{OliveGreen}{\hdashrule[0.5ex]{1.5em}{2pt}{1pt 1pt}} $\kappa_2(x)$
        \end{tabular}
    \end{minipage}

    \vspace{0.4cm}

    \begin{subfigure}[t]{0.48\textwidth}
        \centering
        \includegraphics[width=\linewidth]{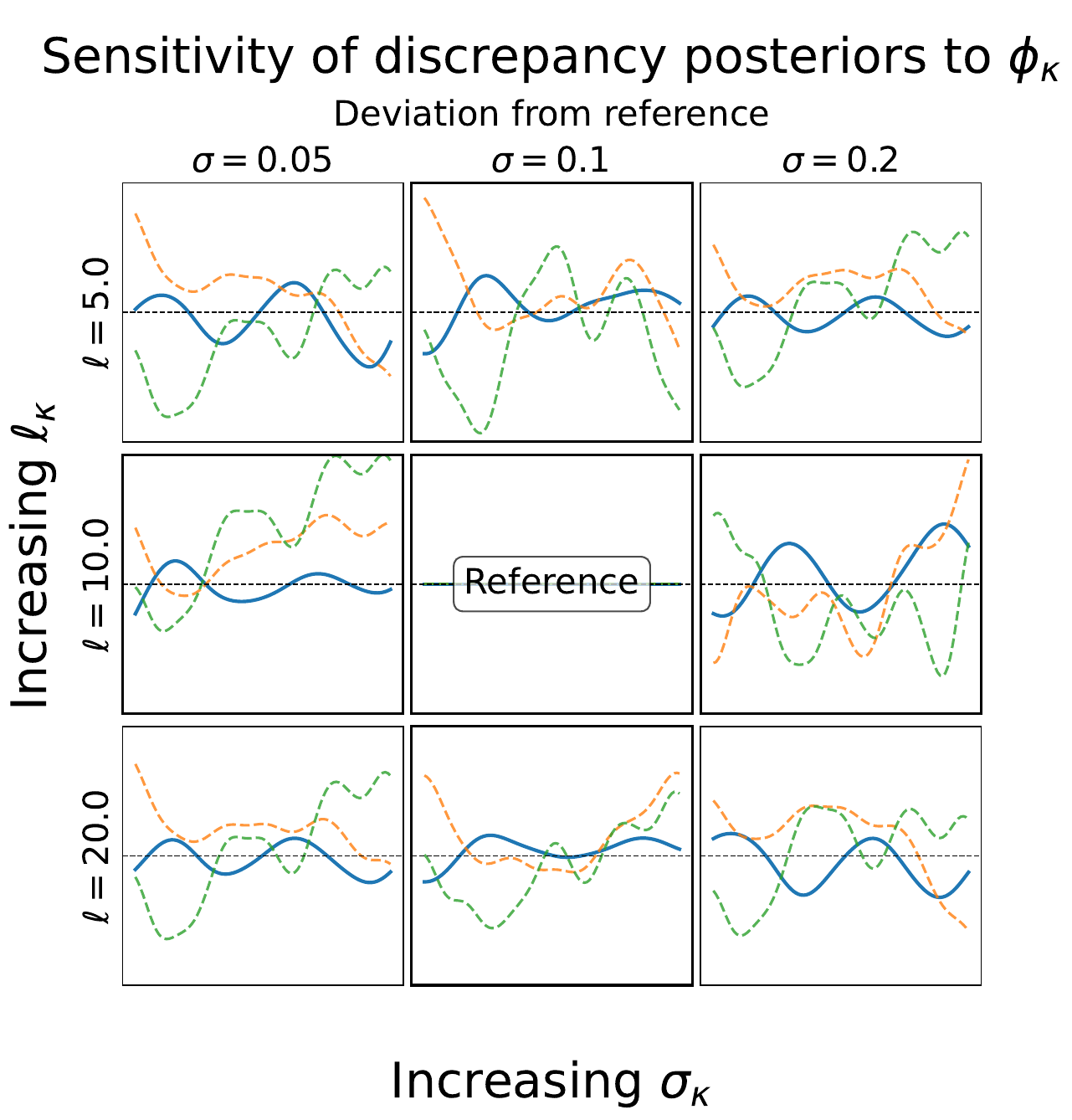}
        \caption{Sensitivity to $\phi_{\kappa}$ for Study No. 1}
        \label{fig:sensitivity_k1}
    \end{subfigure}
    \hfill
    \begin{subfigure}[t]{0.48\textwidth}
        \centering
        \includegraphics[width=\linewidth]{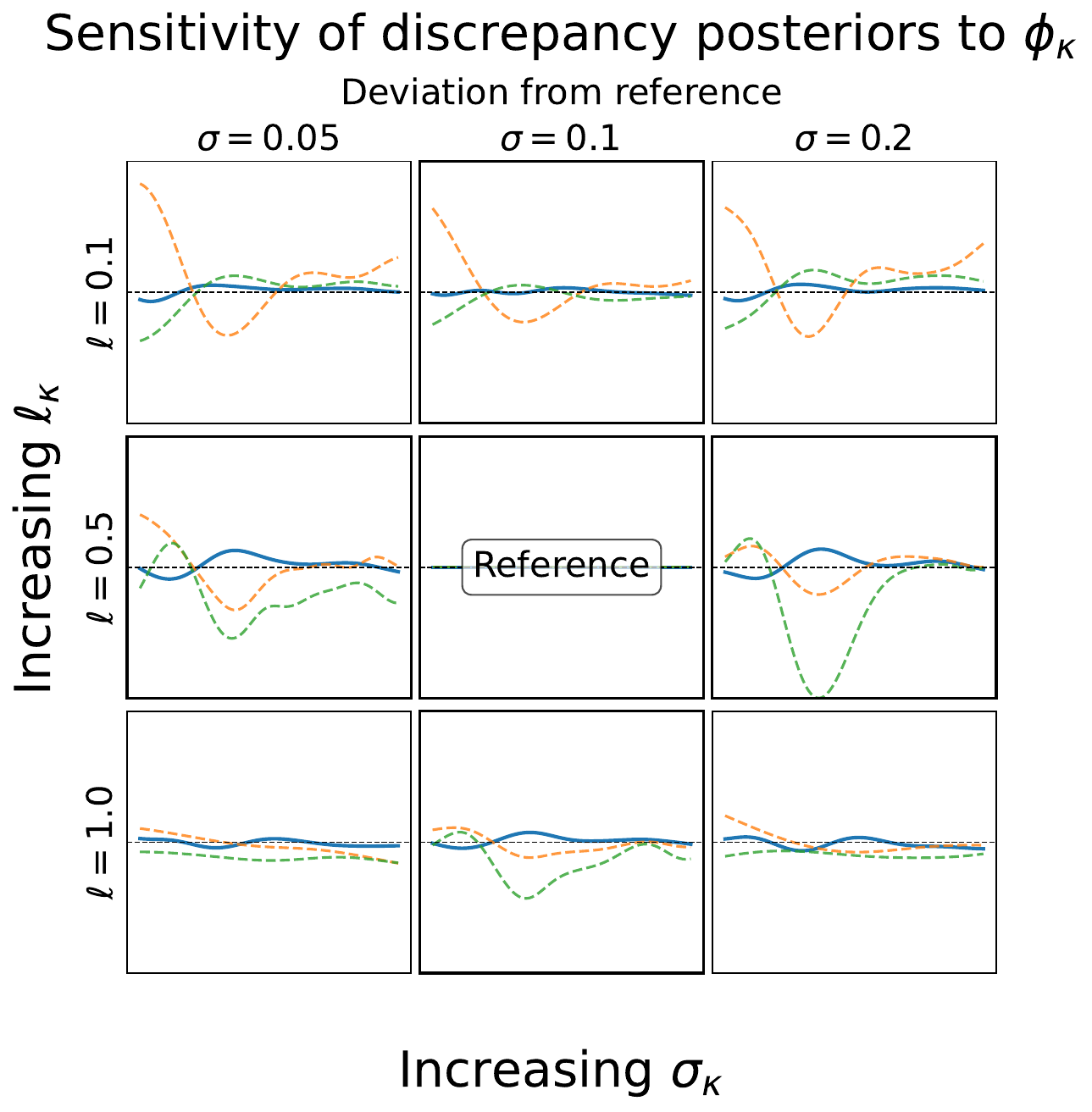}
        \caption{Sensitivity to $\phi_{\kappa}$ for Study No. 2}
        \label{fig:sensitivity_k2}
    \end{subfigure}

    \caption{
    Sensitivity analysis of internal GP models to hyperparameters
    $\{\phi_\kappa,\phi_{\delta_\eta}\}$.
    The left column corresponds to Study No.~1 and the right column
    corresponds to Study No.~2.
    }
    \label{fig:sensitivity_analysis}
\end{figure}


Using the defined loss, one can systematically quantify the quality of model predictions for cross-validation studies. Optimization routines were set up using Optuna \cite{optuna} to minimize loss as a function of applied shrinkage hyperpriors; however, optimal results are not shown here to emphasize the method's generalizability. For the results shown in Section \ref{sec:results}, the recorded NRMSE of each posterior predictive mean with accompanying discrepancy fields is provided in Table \ref{tab:rmse_results}. The results of IBFU are compared against IB and KOH, although only IBFU is capable of decoupling parameter correction error from additive discrepancy error. 

Figure \ref{fig:sensitivity_analysis} demonstrates how modifying the shrinkage priors $\phi_{\delta_\eta}$ and $\phi_\kappa$ modifies the resultant $\delta_\eta(x)$ and $\kappa(x)$ fields. Changes in the $\delta_\eta(x)$ field as a function of the variance hyperprior are evident with $O(0.1)$ modification, while modifications to $\ell$ are less pronounced, requiring O$(10)$ modification to demonstrate significantly different results for both $\phi_{\delta_\eta}$ and $\phi_\kappa$. 

\begin{table}[h!]
\centering
\renewcommand{\arraystretch}{1.2}
\setlength{\tabcolsep}{4pt}

\begin{tabular}{
|C{1.8cm}
|C{2.3cm}
||C{1.8cm}
|C{2.0cm}
|C{2.3cm}|
}
\hline


\multicolumn{2}{|c||}{} 
& $\mathrm{NRMSE}(y)$
& $\mathrm{NRMSE}(\kappa)$
& $\mathrm{NRMSE}(\delta_\eta)$ \\
\hline


\multirow{2}{*}{IBFU}
& Study No. 1
& 0.05
& 1.02
& 1.3 \\
\cline{2-5}

& Study No. 2
& 0.31
& 0.43
& 1.2 \\
\hline


\multirow{2}{*}{IB}
& Study No. 1
& 0.07
& 3.11
& -- \\
\cline{2-5}

& Study No. 2
& 0.06
& 0.71
& -- \\
\hline


\multirow{2}{*}{KOH}
& Study No. 1
& $\approx0^*$
& --
& 0.2 \\
\cline{2-5}

& Study No. 2
& $\approx0^*$
& --
& 0.53 \\
\hline

\end{tabular}

\footnotesize
$*$  lack of error due to overfitting

\caption{
Comparison of normalized root-mean squared error (NRMSE) metrics across calibration methodologies
and conceptual studies. For $\kappa(x)$ corresponding to multiple $\theta$ parameters, the maximum NRMSE corresponding to the most uncertain parameter is reported.
}

\label{tab:rmse_results}

\end{table}

The implemented orthogonalization ensures that each source of model form error can only be attributed to the input parameters which the model is appropriately sensitive to, with all residual error being attributed to additive model discrepancy.  While the current method provides a significant improvement to Bayesian calibration methods, the relative errors found in the cross-validation analysis demonstrate that the identifiability issue is yet to be wholly eliminated.

\subsection{Limitations of IBFU}
\label{subsec:limits}

While IBFU is conveniently applicable to a wide variety of problems with input parameters that are both constant and functional across the domain, the additive discrepancy, as currently implemented, is a zero-mean GP which is better suited to capturing additive discrepancy signals which fluctuate about zero instead of strictly positive or negative predictions. It is also worth noting that if IBFU is able to calibrate the reality response using $\kappa(x)$ fields, this limitation does not necessarily hold. The results of Figure \ref{fig:ddd-md} demonstrate a DDD simulation which consistently underpredicts MD observations; however, because the method found DDD predictions of $\tau^{CRSS}$ to be sensitive to the shear modulus $\mu$, the posterior predictive measures compare well against observation data with reasonable parameter and additive discrepancy fields. 
Another critical assumption being made is the reliability of the prescribed $\theta^0$. In the event expert opinions deviate significantly from observed data, the anchor for parameter corrections may no longer be a strong anchor for updating posterior predictions and one ought to employ loose shrinkage on $\ell_\kappa$.

Evaluating the limiting case of $N_{obs}\rightarrow0$ also illuminates one of the guiding assumptions of IBFU, especially when equipped with conservative shrinkage hyperpriors. In the vanishing data limit, KOH posterior predictions of $\theta$ collapse to the prior, and the joint inference of $\theta$ and $\delta_\eta$ is the cause of the identifiability issue. In the vanishing data limit of the integrated bias with full uncertainty methodology, the posterior covariance again converges with prior covariance, but the same is not necessarily true when shrinkage priors are applied. 
Introducing shrinkage hyperpriors transforms the full discrepancy formulation from a conditionally Gaussian inverse problem into a hierarchical covariance inference problem. In this setting, posterior inference must simultaneously determine the latent discrepancy realizations, the covariance geometries governing their smoothness and amplitudes, and the decomposition of residual structure between embedded and additive discrepancy mechanisms. Consequently, the low-data asymptotic regime becomes dominated not merely by fixed prior covariance assumptions, but by uncertainty over the covariance operators themselves, yielding a non-Gaussian hierarchical posterior governed primarily by the prescribed shrinkage structure. This differs philosophically from KOH, and cannot be evaluated objectively. Orthogonalization further complexifies the evaluation of asymptotic cases, as discussed in Appendix \ref{appendix:limCases}.

\section{Conclusions}
\label{sec:conclusion}

We have demonstrated a novel reimagining of the KOH formalism which is capable of producing predictions  for a constant and functional set of input parameters across the domain. IBFU anchors calibration results about best estimates, which preserves physical information throughout the posterior inference. The principle of orthogonality ensures the various sources of model-form error are attributed to the appropriate fields, whether that is input parameter uncertainty or model-form error to be accounted for via additive discrepancy. Each source of discrepancy has its own model flexibility that can be tuned through careful manipulation of hyperpriors $\phi_\kappa$ and $\phi_{\delta_\eta}$, and guidance on prescription of those hyperparameters is provided. While the identifiability issue attributed with Bayesian methods is not entirely eliminated, the current methodology is a significant improvement over previously developed calibration routines. 

\section*{Acknowledgments}
The author wants to acknowledge the intellectual contributions of Jacob Jeffries, who derived the equivalence between Equations \ref{eq:KOH} the IB method in the limit $||\delta_\theta(x_i)||<\varphi$ for some arbitrarily small $\varphi$. 

This work was supported, in part, at Clemson University by the State of South Carolina through funding for the Battelle Savannah River Alliance Workforce Development Program.


\section{Appendices}\label{}


\appendix
\renewcommand{\theequation}{A\arabic{equation}}
\setcounter{equation}{0}
\input{appendixA}

\bibliographystyle{unsrtnat}
\bibliography{Box_refs}



\end{document}

%% file: appendixA.tex
\section{Analytical Derivations of the Limiting Case $N_{\mathrm{obs}}\rightarrow0$}
\label{appendix:limCases}

\subsection{Integrated Bias Without Explicit Discrepancy}
\label{app:ib_no_delta}

Consider the simplified integrated bias model

\begin{equation}
    y = (\theta_0 + \kappa) + \varepsilon
\end{equation}

where $\theta_0$ is treated as a deterministic calibration parameter and

\begin{equation}
    \kappa \sim \mathcal{N}(0,K_\kappa).
\end{equation}

Assuming independent Gaussian observational noise,

\begin{equation}
    \varepsilon \sim \mathcal{N}(0,\sigma^2 I),
\end{equation}

the conditional distribution of the observations becomes

\begin{equation}
    y \mid \theta_0
    \sim
    \mathcal{N}
    \left(
        \theta_0,
        K_\kappa + \sigma^2 I
    \right).
\end{equation}

The posterior distribution for $\kappa$ is therefore

\begin{equation}
    \kappa \mid y,\theta_0
    \sim
    \mathcal{N}(\mu_\kappa,\Sigma_\kappa),
\end{equation}

with posterior mean

\begin{equation}
    \mu_\kappa
    =
    K_\kappa
    \left(
        K_\kappa + \sigma^2 I
    \right)^{-1}
    (y - \theta_0 \mathbf{1}),
\end{equation}

and posterior covariance

\begin{equation}
    \Sigma_\kappa
    =
    K_\kappa
    -
    K_\kappa
    \left(
        K_\kappa + \sigma^2 I
    \right)^{-1}
    K_\kappa.
\end{equation}

As $N_{\mathrm{obs}}\rightarrow0$, observational information vanishes such that

\begin{equation}
    y - \theta_0 \mathbf{1}
    \rightarrow 0.
\end{equation}

Consequently,

\begin{align}
    \mu_\kappa &\rightarrow 0, \\
    \Sigma_\kappa &\rightarrow K_\kappa.
\end{align}

Therefore, the posterior collapses to the Gaussian process prior,

\begin{equation}
    p(\kappa \mid y,\theta_0)
    \rightarrow
    p(\kappa).
\end{equation}

This implies that, in the low-data limit, inference over the embedded discrepancy field becomes entirely prior driven.

\clearpage

\subsection{Integrated Bias with Full Discrepancy}
\label{app:ib_full_delta}

Now consider the full uncertainty model,

\begin{equation}
    y = (\theta_0 + \kappa) + \delta + \varepsilon.
\end{equation}

Under the assumptions

\begin{align}
    \kappa &\sim \mathcal{N}(0,K_\kappa), \\
    \delta &\sim \mathcal{N}(0,K_\delta),
\end{align}

the marginal observation model becomes

\begin{equation}
    y \mid \theta_0
    \sim
    \mathcal{N}
    \left(
        \theta_0,
        K_\kappa + K_\delta + \sigma^2 I
    \right).
\end{equation}

Define the latent vector

\begin{equation}
    z =
    \begin{bmatrix}
        \kappa \\
        \delta
    \end{bmatrix},
\end{equation}

with prior covariance

\begin{equation}
    \Sigma_0
    =
    \begin{bmatrix}
        K_\kappa & 0 \\
        0 & K_\delta
    \end{bmatrix}.
\end{equation}

Also define

\begin{equation}
    H =
    \begin{bmatrix}
        I & I
    \end{bmatrix}.
\end{equation}

The posterior distribution is then

\begin{equation}
    z \mid y,\theta_0
    \sim
    \mathcal{N}(\mu_z,\Sigma_z),
\end{equation}

where

\begin{equation}
    \mu_z
    =
    \Sigma_0 H^T
    \left(
        H\Sigma_0H^T + \sigma^2 I
    \right)^{-1}
    (y-\theta_0\mathbf{1}).
\end{equation}

The posterior expectations of $\kappa$ and $\delta$ become

\begin{align}
    \mathbb{E}[\kappa \mid y]
    &=
    K_\kappa
    \left(
        K_\kappa + K_\delta + \sigma^2 I
    \right)^{-1}
    (y-\theta_0\mathbf{1}),
    \\
    \mathbb{E}[\delta \mid y]
    &=
    K_\delta
    \left(
        K_\kappa + K_\delta + \sigma^2 I
    \right)^{-1}
    (y-\theta_0\mathbf{1}).
\end{align}

In the low-data limit,

\begin{equation}
    y-\theta_0\mathbf{1}\rightarrow0,
\end{equation}

which implies

\begin{align}
    \mathbb{E}[\kappa \mid y] &\rightarrow 0, \\
    \mathbb{E}[\delta \mid y] &\rightarrow 0.
\end{align}

Therefore,

\begin{equation}
    p(\kappa,\delta \mid y)
    \rightarrow
    p(\kappa,\delta),
\end{equation}

indicating that the decomposition becomes entirely governed by prior covariance assumptions.

\clearpage

\subsection{Full Discrepancy with Shrinkage Hyperpriors}
\label{app:ib_shrinkage}

The previous derivation assumes fixed covariance operators for both discrepancy fields. Consider now covariance kernels parameterized by hyperparameters,

\begin{align}
    K_\kappa &= K_\kappa(\ell_\kappa,\sigma_\kappa^2), \\
    K_\delta &= K_\delta(\ell_\delta,\sigma_\delta^2),
\end{align}

where the correlation length scales and variances are themselves random variables,

\begin{align}
    (\ell_\kappa,\sigma_\kappa^2)
    &\sim
    p(\ell_\kappa,\sigma_\kappa^2),
    \\
    (\ell_\delta,\sigma_\delta^2)
    &\sim
    p(\ell_\delta,\sigma_\delta^2).
\end{align}

Define the collective hyperparameter vector

\begin{equation}
    \phi
    =
    (
        \ell_\kappa,
        \sigma_\kappa^2,
        \ell_\delta,
        \sigma_\delta^2
    ).
\end{equation}

Conditioned on $\phi$, the posterior remains Gaussian. However, marginalizing over $\phi$ yields

\begin{equation}
    p(\kappa,\delta \mid y)
    =
    \int
    p(\kappa,\delta \mid y,\phi)
    p(\phi \mid y)
    \, d\phi,
\end{equation}

which is a continuous mixture of Gaussian distributions.

Consequently, the posterior is no longer jointly Gaussian and may become multimodal. No closed-form marginal posterior generally exists.

In the limiting case $N_{\mathrm{obs}}\rightarrow0$, the effective covariance becomes entirely governed by the hyperprior structure,

\begin{equation}
    \Sigma \rightarrow K(\phi),
\end{equation}

such that inference transitions from a standard Gaussian conditioning problem to a hierarchical covariance inference problem dominated by hyperprior assumptions.